\documentclass[aps, twocolumn, pra, superscriptaddress, amsmath, showpacs, tightenlines, letterpaper] {revtex4}
\usepackage{epsfig,graphicx,times}
\usepackage{amstext}
\usepackage{amsmath}            
\usepackage{amssymb}            
\usepackage{amsthm}
\usepackage{leftidx}
\usepackage{mathrsfs}
\usepackage{graphicx}           
\usepackage{subfigure}
\usepackage{latexsym}
\usepackage{bm}
\usepackage{epstopdf}
\usepackage{epsfig}

\begin{document}

\title{Quantum Simulation of Pairing Hamiltonians with Nearest-Neighbor Interacting Qubits}

\author{Zhixin Wang}
\affiliation{Department of Microelectronics and Nanoelectronics,
Tsinghua University, Beijing 100084, China}

\author{Xiu Gu}
\affiliation{Department of Microelectronics and Nanoelectronics, Tsinghua University, Beijing 100084, China}
\affiliation{Institute of Microelectronics, Tsinghua University,
Beijing 100084, China}

\author{Lian-Ao Wu}
\affiliation{Ikerbasque, Basque Foundation for Science, 48011 Bilbao and Department of Theoretical Physics and History of Science,
The Basque Country University (UPV/EHU), PO Box 644, 48080 Bilbao, Spain}

\author{Yu-xi Liu}
\email{yuxiliu@mail.tsinghua.edu.cn}
\affiliation{Department of Microelectronics and Nanoelectronics, Tsinghua University, Beijing 100084, China}
\affiliation{Institute of Microelectronics, Tsinghua University,
Beijing 100084, China}
\affiliation{Tsinghua National Laboratory for Information Science and Technology (TNList), Beijing 100084, China}

\date{\today}

\begin{abstract}
  Although a universal quantum computer is still far from reach,  the tremendous advances in controllable quantum devices, in particular with solid-state systems, make it possible to physically implement ``quantum simulators". Quantum simulators are physical setups able to simulate other quantum systems efficiently that are intractable on classical computers. Based on solid-state qubit systems with various types of nearest-neighbor interactions, we propose a complete set of algorithms for simulating pairing Hamiltonians. Fidelity of the target states corresponding to each algorithm is numerically studied. We also compare algorithms designed for different types of experimentally available Hamiltonians and analyze their complexity. Furthermore, we design a measurement scheme to extract energy spectra from the simulators. Our simulation algorithms might be feasible with state-of-the-art technology in solid-state quantum devices.
\end{abstract}


\maketitle

\pagenumbering{arabic}

\section{Introduction}

Classical computers fail to efficiently simulate quantum systems with complex many-body interactions due to the exponential growth of variables for characterizing these systems~\cite{Buluta09,BulutaRMP}. For instance, $2^N$ parameters are required for the complete description of a quantum system composed of $N$ entangled spin-1/2 particles. In the 1980s, quantum simulation was proposed to solve such an exponential explosion problem using a controllable quantum system ~\cite{Feynman82}. In 1996, it was shown that a quantum computer only containing few-particle interactions can be used to efficiently simulate many-body quantum Hamiltonians~\cite{Lloyd96}. Recently quantum simulation has attracted extensive attention in condensed matter physics~\cite{Britton12, Simon11}, high energy physics~\cite{Gerritsma10} and quantum chemistry~\cite{Lanyon10} due to the rapid progress on coherent control of quantum systems aimed at quantum information processing~\cite{Monroe13,Awschalom13,Blatt10,Bloch10,Hanson10}. Quantum simulators using trapped ions~\cite{Blatt12,Barreiro11,Lanyon11,Kim10}, cold atoms~\cite{Bloch12,Struck11} and photons~\cite{Aspuru-Guzik12} have already been experimentally demonstrated  to some extent.

Quantum simulators are classified into analog and digital ones~\cite{Buluta09} . An analog quantum simulator is a controllable quantum system mimicking the behaviors of the target quantum system whose evolution can be effectively mapped onto the simulator~\cite{Simon11}, while a digital quantum simulator normally imitates the time evolution operator of the target system through the implementation of a series of elementary quantum gates~\cite{Lanyon11}. Practically, these two approaches are often combined. A simulation task can be performed through the free evolution of quantum simulators combined with external logic gates at given time instants. Quantum simulators containing tens of qubits are practically useful to carry out classically intractable tasks while the number of qubits required for practical quantum computing is much larger~\cite{Buluta09}. Therefore, in comparison with universal quantum computing, quantum simulation is more feasible in the near future, and the simulation algorithms for certain task using existed Hamiltonians are strongly desired.

Pairing Hamiltonians, for example, BCS Hamiltonian in conventional superconductors, feature long-range many-body interactions which are generally intractable on classical computers. Nevertheless, large-scale numerical calculations based on pairing Hamiltonians are of great importance, for instance in mesoscopic condensed matter, ultrasmall metallic grains and heavy nuclei~\cite{Wu02}. To tackle this problem, a polynomial-time quantum algorithm based on a nuclear magnetic resonance (NMR) quantum computer was proposed~\cite{Wu02}, and has been demonstrated experimentally~\cite{Brown2006,Yang06}. However, liquid NMR has several constrains that make NMR quantum computer not scalable~\cite{Ladd10}. Therefore, the large-scale implementation of the NMR-based quantum algorithm is unlikely with the state-of-the-art technology.

Rapid progress of superconducting quantum circuits has been witnessed in the past decades~\cite{Ladd10, Devoret13, Clarke08, You05, You11}. This enables them to be one of the most promising candidates towards practical quantum information processing. Based on nearest-neighbor coupled superconducting qubits, single-qubit and two-qubit gates with fidelity at the threshold of fault tolerant quantum computing has been realized recently~\cite{Barends14}. Various theoretical and experimental explorations on quantum simulation have been carried out using this approach~\cite{Houck12, Underwood12, Roushan14}. The unique flexibility on design and fabrication of superconducting circuits enables wide tunability in extensive Hamiltonian parameter ranges and the techniques for scaling up are compatible with those for modern integrated circuits. Both of the above aspects are significant advantages for superconducting quantum circuits to serve as practical quantum simulators. Moreover, the research on other solid-state qubit devices, e.g. quantum dot in semiconductors~\cite{Eriksson-MRS-2013} and defect systems~\cite{Gordon-MRS-2013}, has also made significant progress in past years. Therefore, it would be of great interest to update the algorithm in NMR system~\cite{Wu02} to simulate the paring Hamiltonian using solid-state qubit systems.

Here we propose a complete set of simulation algorithms and a measurement scheme to simulate many-body pairing models using various Hamiltonians existed in solid-state qubit systems. The algorithms are suitable for a wide range of quantum systems, especially superconducting quantum circuits and semiconducting qubit systems. Sec.~\ref{tool_task} is a brief description of our simulation task and available theoretical Hamiltonians for quantum simulation. General simulation algorithms based on qubits with various types of nearest-neighbor interactions are presented in Sec.~\ref{algo1}. The fidelity of the algorithms and its variation with various parameters are numerically studied in Sec.~\ref{numerical}. The complexity of the algorithms as well as its parameter dependence is analyzed in Sec. \ref{complex}.
We have also designed an effective measurement scheme based on the entanglement of a single qubit in the simulator and an ancillary qubit in Sec.~\ref{section_mea}, through which crucial information in the energy spectrum could be extracted. We finally summarize our results in Sec.~\ref{conclusion}.

\section{Preliminaries} \label{tool_task}

For the completeness of this paper, first we briefly review pairing Hamiltonians and outline our simulation tasks.

\subsection{Pairing Hamiltonians and qubit representation} \label{Pauli}

The general BCS pairing Hamiltonian is extensively used in condensed matter physics and nuclear physics, and has the form:
\begin{equation} \label{BCS}
H_{\textrm{BCS}}=\sum_{m=1}^N\frac{\epsilon_m}{2}(n_m^F+n_{-m}^F) + \sum_{l=1}^N \sum_{m=1}^N V_{ml} c_m^\dagger c_{-m}^\dagger c_{-l}c_l,
\end{equation}
where $c_{\pm m}^\dagger$ and $c_{\pm l}$ are fermionic creation and annihilation operators, and $n_{\pm m}^F = c_{\pm m}^\dagger c_{\pm m}$ are fermionic number operators.
As has been analyzed (see for example Ref.~\cite{Wu02}), the BCS pairing Hamiltonian made by fermionic pair operators can be mapped onto qubit operators $\sigma_m^x$, $\sigma_m^y$ and $\sigma_m^z$ through the transformation $ \{ \sigma_m^x, \sigma_m^y, \sigma_m^z \} = \{ c_m^\dagger c_{-m}^\dagger + c_{-m} c_m  , i c_{-m} c_m - i c_m^\dagger c_{-m}^\dagger, n_{m}^F + n_{-m}^F - 1 \}$. With the mapping, we can rewrite the Hamiltonian in Eq. (\ref{BCS}) as,
\begin{equation} \label{pair}
H_{p}=\sum_{m=1}^N\frac{\varepsilon_m}{2}\sigma_m^z + \sum_{m<l} \frac{V_{ml}}{2} \left(\sigma_m^x\sigma_{l}^x + \sigma_m^y\sigma_{l}^y\right),
\end{equation}
with $\varepsilon_m = \epsilon_m + V_{mm}$.



\begingroup
\begin{table}[ptb]
\caption{Summary for various interaction Hamiltonians in solid-state systems with single-qubit Hamiltonians $H_{0}=\sum_{l=1}^N (\omega_l\sigma_l^z/2)$.  Note that $J_{l}^{x}=J_{l}^{y}=J_{l}$ for the Heisenberg model in the table.}
\label{tab1}
\begin{ruledtabular}
\begin{tabular}{c|p{8cm}}
  Interaction types& Interaction Hamiltonians\\
\hline
 Longitudinal Ising model& $ H_{\text{Ising,L} }=H_{0}+ \sum_{l=1}^{N-1}J_l\sigma_l^z\sigma_{l+1}^z  $ \\
\hline   Transverse Ising model & $ H_{\text{Ising,T} }=H_{0}+ \sum_{l=1}^{N-1}J_l\sigma_l^x\sigma_{l+1}^x  $\\
\hline XY model& $H_{\textrm{XY}} =H_{0}+\sum_{i=x,y} \sum_{l=1}^{N-1}J_l\sigma_l^i\sigma_{l+1}^i  $\\
\hline Heisenberg model& $H_{\textrm{H}} =H_{0}+ \sum_{i=x,y,z}\sum_{l=1}^{N-1}J_l^i \sigma_l^i\sigma_{l+1}^i  $
\end{tabular}
\end{ruledtabular}\label{table1}\end{table}\endgroup

\subsection{Qubits with Nearest-Neighbor Coupling}

Solid-state qubits can be coupled through various types of interactions. For instance, superconducting qubits can be coupled to their nearest neighbors through capacitances~\cite{Yamamoto03}, inductances~\cite{Majer05} or Josephson junctions~\cite{Siewert00}. Different interaction models~\cite{Schuch03} resulted from different coupling schemes can be classified into four categories of commonly used interaction Hamiltonians: longitudinal Ising types~\cite{Makhlin99, Orlando99,You2002}, transverse Ising types~\cite{Pashkin03, McDermott05, Bertet06}, XY types~\cite{Imamoglu99, Mozyrsky01,Quiroga1999} and Heisenberg types~\cite{Loss98,Kane98,Burkard99,Hu2000,Vrijen00}. These four different types of nearest-neighbor Hamiltonians can be unified as
\begin{equation}\label{nearest}
H = H_0 + H_I
\end{equation}
 with $H_0$ denoting the single-qubit Hamiltonian
\begin{equation}\label{eq:4}
H_{0} = \sum_{l=1}^N\frac{1}{2} \omega_l \sigma_l^z
\end{equation}
and $H_I$ denoting the interaction Hamiltonian
\begin{equation}
H_{I} = \sum_{l=1}^{N-1} (J_l^x\sigma_l^x\sigma_{l+1}^x +J_l^y\sigma_l^y\sigma_{l+1}^y+J_l^z\sigma_l^z\sigma_{l+1}^z ).
\end{equation}
Here $\sigma_l^x, \sigma_l^y, \sigma_l^z$ are Pauli matrices in the basis of $\sigma_{l}^{z}$, and $l$ denotes the $l$th qubit.  Parameters $J_l$ $(l = 1, \dots, N-1)$ denote the coupling strength between the $l$th and $(l+1)$th qubits.

Each of these four types of Hamiltonians is a special case of Eq.~(\ref{nearest}) with parameters being properly chosen. The Hamiltonian~(\ref{nearest}) can be thus reduced to (i) longitudinal Ising Hamiltonian for parameters $J_l^x = J_l^y = 0$  and $J_l^z = J_l$; (ii)  the transverse Ising Hamiltonian for parameters $J_l^y = J_l^z = 0$ and $ J_l^x = J_l$; (iii) the XY Hamiltonian for parameters $J_l^z = 0$ and $J_{l}^{x}=J_{l}^{y}=J_{l}$; (iv) the Heisenberg Hamiltonian for parameters $J_l^x = J_l^x = J_l^y = J_l$. For clarity, we summarize them in Table~\ref{table1}.

Assuming that qubit systems with nearest-neighbor couplings in Eq.~(\ref{nearest}) are available experimentally, we can hence use them to simulate dynamical behaviors of pairing Hamiltonians in Eq.~(\ref{pair}) with the help of single-qubit operations. These operations will be done by applying external fields $F=\sum_{l=1}^N ( f_l^x\sigma_l^x + f_l^y\sigma_l^y + f_l^z\sigma_l^z ) $ to individual qubits.  It is clear that the pairing models in Eq.~(\ref{pair}) do not share the same form of Hamiltonians with in Eq.~(\ref{nearest}). Therefore,  it is necessary to design algorithms to simulate these pairing Hamiltonians using the four types of interaction Hamiltonians mentioned above.

\subsection{Quantum simulations}

Our goal is to simulate the BCS-type pairing Hamiltonian in Eq.~(\ref{pair}) which can be rewritten as
\begin{equation} \label{BCS1}
H_{p} = H_{p0} + H_{pI},
\end{equation}
where ${H_{p0}} = \sum_{m=1}^N \varepsilon_m \sigma_m^z / 2$ is the single-qubit Hamiltonian and ${H_{pI}} = \sum_{l>m}^N V_{ml} (\sigma_m^x\sigma_{l}^x +\sigma_m^y\sigma_{l}^y) / 2$ is the interaction Hamiltonian.

We now come to simulate parameters $\varepsilon_m$ $(m = 1 \dots N)$ and $V_{ml}$ $(m,l = 1 \dots N, m<l)$ in Eq.~(\ref{BCS1}) with each of the Hamiltonians in Table~\ref{table1} 
by using the decoupling and recoupling techniques in Ref. ~\cite{Leung00}.  The single-qubit terms ${\varepsilon_m}\sigma_m^z /2$ $(m = 1, \ldots, N)$ and two-qubit interaction Hamiltonian terms $V_{ml} \left(\sigma_m^x\sigma_{l}^x +\sigma_m^y\sigma_{l}^y\right) / 2$ $(l,m = 1, \ldots, N, l<m)$ will be simulated separately through dynamical evolution. Special care must be taken when the separated terms are put all together by using Trotter's formula since operators $\sigma_m^x, \sigma_m^y$,  and $\sigma_m^z$ do not commute with each other. Another challenge is to simulate long-range interaction terms in $H_p$ through simulators only containing nearest neighbor interactions. Therefore, methods for extending the range of interaction must be properly designed.

We note that the tunability of parameters $\omega_l$ $(l = 1, \dots, N)$ and $J_l^i$ $(i=x, y, z; l = 1, \dots, N-1)$ affects the efficiency of the algorithms. Algorithms for simulators with constant parameters presented in the following section are often more complex. Simplification is allowed if some of the parameters are tunable during the simulation process. A comparison of fidelity and complexity between simulators with constant and tunable coupling parameters will be given in Sec. \ref{numerical} and Sec. \ref{complex}.

\section{Simulation Algorithms} \label{algo1}

In this section, we will give detailed discussion on how to simulate pairing Hamiltonian using Ising-type, or XY-type, or Heisenberg-type interaction Hamiltonians as listed in Table~\ref{table1}.

\subsection{Algorithm for Simulators with Longitudinal Ising Hamiltonian} \label{solu_LIsing}

We first study an algorithm to simulation pairing Hamiltonians using longitudinal Ising Hamiltonian. The simulation algorithm needs two steps~\cite{Wu02}. The first step is to simulate $H_{p0}$ and the second one is to mimic $H_{pI}$. We then combine $H_{p0}$ and $H_{pI}$ to obtain the complete pairing Hamiltonian. The detailed description for these steps is given as below.

\subsubsection{Simulation Algorithm for Individual Terms $H_{p0}$} \label{solu_LIsing_H0}

\begin{figure}[htb]
 \centerline{\includegraphics[width = 8.5 cm]{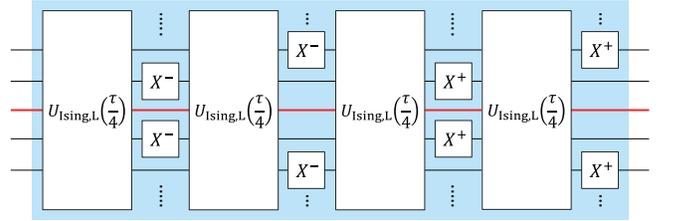}}
 \caption{The quantum circuit to realize $U_l^z(\tau)$ from $U_{\textrm{Ising,L}}$. The red bold line stands for the $l$th qubit in the system, whose individual evolution term is going to be extracted through this step. Black dots on both sides stand for the periodic extension of the logic gates to the rest of the system in the given direction. Here, the period is two. $X^\pm$ stand for external gates $e ^ {\pm i\frac{\pi}{2} \sigma^x}$.}
 \label{fig1}
\end{figure}

We now use the simulators with longitudinal Ising-type interaction $H_{\rm Ising, L}$ in Table~\ref{table1} to simulate ${H_{p0}}$.  The operator
\begin{equation}
U_{\textrm{Ising,L}}(\tau) = e ^ {-i\tau H_{\textrm{Ising,L}}}
\end{equation}
denotes the time evolution operator of the quantum simulator. Let us consider the time evolution operator $U_l^z(\tau) = \exp (-i\tau \omega_l \sigma_l^z / 2 )$ of the $l$th qubit. We show that $U_l^z(\tau)$ can be given as
\begin{equation}\label{fU{z+zz}--U{l,z}}
\begin{split}
&U_l^z(\tau) = \\
\;&\left( \bigotimes_{j'' \neq l} e ^ {i\frac{\pi}{2} \sigma_{j''}^x} \right) U_{\textrm{Ising,L}} \left( \frac{\tau}{4} \right) \left( \bigotimes_{j'} e ^ {i\frac{\pi}{2} \sigma_{j'}^x} \right) U_{\textrm{Ising,L}} \left( \frac{\tau}{4} \right) \\
\;&\left( \bigotimes_{j''\neq l} e ^ {-i\frac{\pi}{2} \sigma_{j''}^x} \right) U_{\textrm{Ising,L}} \left( \frac{\tau}{4} \right)
\left( \bigotimes_{j'} e ^ {-i\frac{\pi}{2} \sigma_{j'}^x} \right) U_{\textrm{Ising,L}} \left( \frac{\tau}{4} \right)
\end{split}
\end{equation}
 using $U_{\rm Ising, L}(\tau/4)$. Here $j'$ and $j''$ denote the $j^\prime$th and the $j^{\prime\prime}$th qubits in the simulator.  $j'$ is an even (odd) number and $j''$ is an odd (even) number if $l$ is an odd (even) number, where $j'' \neq l$. Figure~\ref{fig1} shows the corresponding quantum circuits.

For a given $U_l^z(\tau)$, we can extract the time evolution  $H_{\textrm{Ising,L}}$ of each single qubit and then simulate $H_{p0}$ through the circuit $e^{-itH_{p0}} = \bigotimes_{m=1}^N U_m^z (\tau)$ with $\tau=\varepsilon_m t/\omega_m$. Therefore, individual $U_l^z(\tau)$ is the building block when simulating $H_{p0}$. In the following subsections we will show that $U_l^z(\tau)$ can be simulated on simulators with other types of interactions.

\subsubsection{Simulation Algorithm for Interactions $H_{p I}$ and $H_{p}$} \label{solu_LIsing_HI}

Before simulating $H_{pI}$ of the pairing Hamiltonian in Eq.~(\ref{BCS1}) we first consider a time evolution operator $U_{l,m}^{zz}(\tau) = \exp (-i\tau J_l \sigma_l^z \sigma_{m}^z )$. If the simulator has longitudinal Ising nearest neighbor couplings,  $U_{l,l+1}^{zz}(\tau)$
can be directly obtained from $U_{\textrm{Ising,L}}(\tau)$ through the circuit as in Fig.~\ref{U{LIsing}--U{zz,l}}, which can be expressed as
\begin{equation} \label{LI_ll}
\begin{split}
U_{l,l+1}^{zz}(\tau)= \left(\bigotimes_{j'<l}e^{i\frac{\pi}{2} \sigma_{j'}^x}\right)\left(\bigotimes_{j''> l+1} e^{i\frac{\pi}{2} \sigma_{j''}^x}\right) &U_{\textrm{Ising,L}} \left(\frac{\tau}{4}\right) \\
\quad \, \left(\bigotimes_{j''\leq l} e^{i\frac{\pi}{2} \sigma_{j''}^x}\right) \left(\bigotimes_{j'\geq l+1} e^{i\frac{\pi}{2} \sigma_{j'}^x}\right)
 & U_{\textrm{Ising,L}} \left(\frac{\tau}{4}\right) \\
\left(\bigotimes_{j'<l}e^{-i\frac{\pi}{2} \sigma_{j'}^x}\right)\left(\bigotimes_{j''> l+1} e^{-i\frac{\pi}{2} \sigma_{j''}^x}\right) & U_{\textrm{Ising,L}} \left(\frac{\tau}{4}\right) \\
\left(\bigotimes_{j''\leq l} e^{-i\frac{\pi}{2} \sigma_{j''}^x}\right) \left(\bigotimes_{j'\geq l+1} e^{-i\frac{\pi}{2} \sigma_{j'}^x}\right)
 & U_{\textrm{Ising,L}} \left(\frac{\tau}{4}\right).
\end{split}
\end{equation}
Here, $j^{\prime}$ and $j^{\prime\prime}$ for the $j^{\prime}$th and the $j^{\prime\prime}$th qubits in the simulator satisfy the condition that $j'$ is even (odd) and $j''$ is odd (even) if $l$ is odd (even), where $j''$ may equal to $l$. Using the time evolution operator $U_{l,l+1}^{zz}(\tau)$ in Eq.~(\ref{LI_ll})  and single-qubit operations, we have
\begin{eqnarray}
U_{l,l+1}^{xx+yy}(\tau)& =& \exp (-i\tau H_{l}^{xy} )=
e^{i\frac{\pi}{4} Y_{l,l+1}} U_{l,l+1}^{zz}(\tau) e^{-i\frac{\pi}{4} Y_{l,l+1}}\nonumber\\
&\times& e^{i\frac{\pi}{4} X_{l,l+1}}  U_{l,l+1}^{zz}(\tau)  e^{-i\frac{\pi}{4}X_{l,l+1}}
+ O(J^2 \tau^2).\label{eq:10}
\end{eqnarray}
Here the Hamiltonian $H_{l}^{xy}$ denotes the interaction between $l$th and $(l+1)$th qubits in the $xy$ plane, and
\begin{equation}\label{eq:11}
H_{l}^{xy}=J_l ( \sigma_l^x \sigma_{l+1}^x + \sigma_l^y \sigma_{l+1}^y ).
\end{equation}
We hereafter label the sum of Pauli operators of $l$th and $(l+1)$th qubits as
\begin{equation}
P_{l,l+1}=\sigma_{l}^p + \sigma_{l+1}^p
\end{equation}
where $P=X, Y, Z$ correspond to $p=x,y,z$, respectively. For instance, $Y_{l,l+1}$ is expressed as $Y_{l,l+1}=\sigma_{l}^y+\sigma_{l+1}^y$.

\begin{figure}[htb]
  \centerline{
  \includegraphics[width = 8.5 cm]{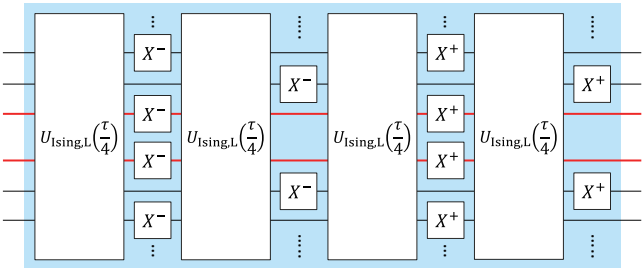}}
  \caption{The quantum circuit for simulating $U_{l,l+1}^{zz}(\tau)$ directly from $U_{\textrm{Ising,L}}(\tau)$. The red bold lines stand for the $l$th and the $(l+1)$th qubits in the system. For extension represented by the black dots on both sides, the period is two.}
  \label{U{LIsing}--U{zz,l}}
\end{figure}

With all nearest-neighbor coupling operators $U_{l,l+1}^{zz}(\tau)$ and $U_{l,l+1}^{xx+yy}(\tau)$ being simulated, we can extend the nearest neighbor interactions to long-range interactions and eventually obtain the total BCS interactions $H_{pI}$ by following the method proposed in Ref.~\cite{Wu02}. When $H_{p0}$ and $H_{pI}$ are both available, the total Hamiltonian can be obtained by Trotter's formula $e^{-itH_p} = e^{-it H_{p0}} e^{-it H_{pI}} + O(t^2)$. In comparison with the exponential complexity to carry out the same task on classical computers, ideally we only needs $O(N^4)$ external single-qubit quantum gates to implement the simulation algorithm.

The discussions above show that the simulations of the nearest-neighbor propagators $U_{l,l+1}^{zz}(\tau)$ and $U_{l,l+1}^{xx+yy}(\tau)$ are crucial in the entire simulating process. In the following subsections, we will explain that $U_{l,l+1}^{zz}(\tau)$ and $U_{l,l+1}^{xx+yy}(\tau)$ can also be simulated through other type of Hamiltonians.

\subsection{Algorithm for Simulators with Heisenberg Nearest-Neighbor Interaction}  \label{solu_Hei}

\subsubsection{Simulation Algorithm for $H_{p0}$}  \label{solu_Hei_H0}

Based on Trotter's formula,  we can decompose the time evolution operator with Heisenberg type interaction Hamiltonian into
\begin{eqnarray}
U_{\textrm{H}}(\tau) &= &e^{-i\tau H_{\textrm{H}}}\approx \exp\left(\sum_{l=1}^N\frac{\omega_{l}^z}{2}\sigma_{z}\right) \exp \left(-i \tau\sum_{l=1}^N H_l^{zz} \right) \nonumber\\
&\times&\exp  \left(-i\tau \sum_{l=1}^{N-1} H_{l}^{xy} \right)+ O(J^2 \tau^2),
\end{eqnarray}
with $H^{zz}_{l}= J_l \sigma_l^z \sigma_{l+1}^z $.

\begin{figure}[htb]
  \centerline{
    \includegraphics[width=4.5cm]{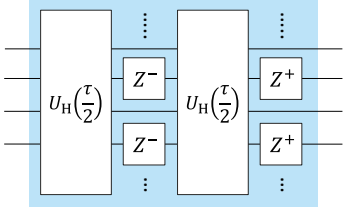}}
  \caption{Quantum circuit to simulate $U_{\textrm{Ising,L}}(\tau)$ from $U_{\textrm{H}}(\tau)$. $Z^\pm$ stand for external gates $e ^ {\pm i\frac{\pi}{2} \sigma^z}$.  Black dots on both sides stand for the periodic extension of the logic gates starting from any single qubit. Here, the period is two.}
  \label{fig3}
\end{figure}

When the quantum simulators possess Heisenberg type interaction, we can first simulate longitudinal Ising Hamiltonian using Heisenberg Hamiltonian and then $H_{p0}$ in terms of longitudinal Ising Hamiltonian. The latter step has already been solved. Assume that error up to  $O(J^2 \tau^2)$ is tolerable, we could obtain the longitudinal Ising Hamiltonian through
\begin{equation} \label{Hei_L2}
\begin{split}
U_{\textrm{Ising,L}}(\tau) \approx &\left(\bigotimes_j e^{i\frac{\pi}{2}\sigma_{2j}^z}\right) U_{\textrm{H}} \left(\frac{\tau}{2}\right) \left(\bigotimes_je^{-i\frac{\pi}{2}\sigma_{2j}^z}\right) U_{\textrm{H}} \left(\frac{\tau}{2}\right),
\end{split}
\end{equation}
Here $2j$ denotes that the qubits with even indices are rotated $\pm \pi/2$ around the $z$ direction, the qubits with odd indices remain unchanged. We can also rotate all qubits with odd indices $\pm\pi/2$ around $z$ axis to obtain the same result. Figure~\ref{fig3} shows the corresponding quantum circuit.

\subsubsection{Simulation Algorithm for Interaction $H_{pI}$}  \label{solu_Hei_HI}

\begin{figure}[htb] 
  \centerline{
  \subfigure[]{
    \includegraphics[width=4.0cm]{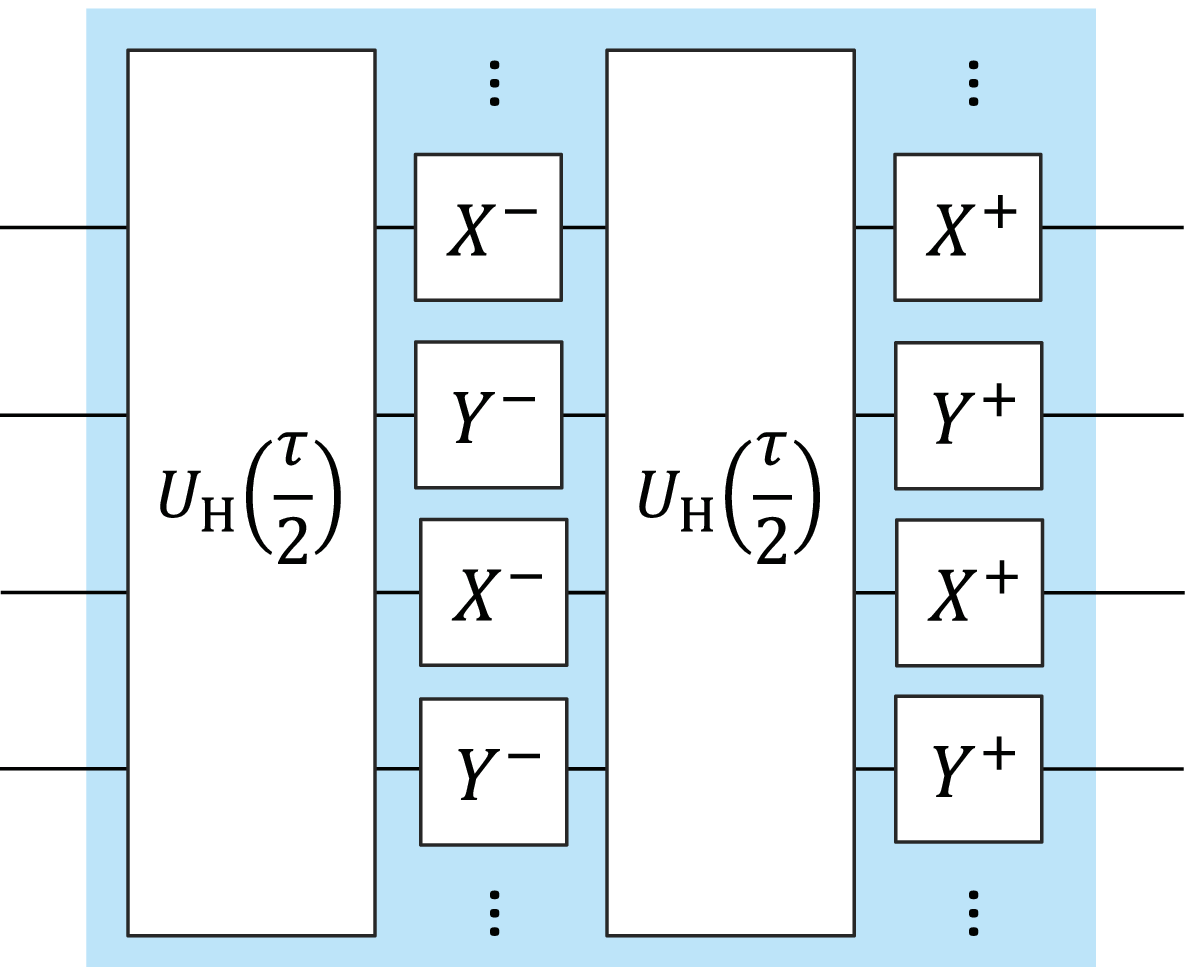}}
  \hspace{0.05cm}
  \subfigure[]{
    \includegraphics[width=4.0cm]{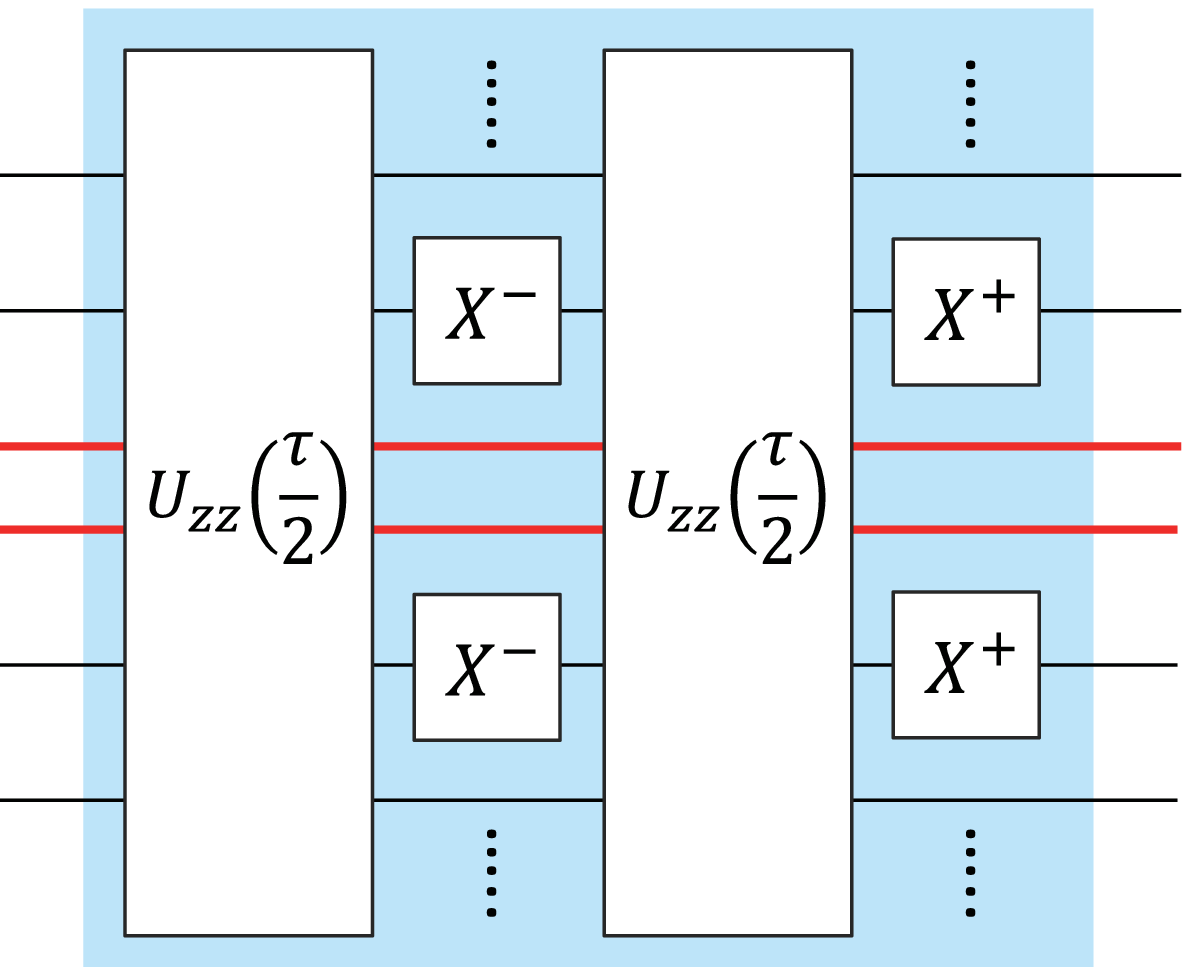}}
    }
  \caption{The quantum circuits to simulate (a) $U_{zz}(\tau)$ from $U_{\textrm{H}}(\tau)$ and (b) $U_{l,l+1}^{zz}(\tau)$ from $U_{zz}(\tau)$. The red bold lines stand for the $l$th and the $(l+1)$th qubits in the system, whose interaction is to be extracted through this step. The period for the extension on both sides represented by black dots is two. $Y^\pm$ stand for external gates $e ^ {\pm i\frac{\pi}{2} \sigma^y}$. }\label{fig4}
\end{figure}

We now design the algorithm to simulate the long range XY interaction terms in the pairing Hamiltonian in Eq.~(\ref{BCS1}) using a quantum simulator with Heisenberg interaction. Let us first consider two time evolution operators
\begin{equation}
U_{zz}(\tau) = \exp (-i\tau \sum_{l=1}^{N-1}J_l\sigma_l^z\sigma_{l+1}^z ),
\end{equation}
and
\begin{equation}
U_{l,m}^{zz}(\tau) = \exp (-i\tau J_l \sigma_l^z \sigma_{m}^z ).
\end{equation}
As shown in Fig.~\ref{fig4}(a), $U_{zz}(\tau)$ can be simulated through $U_{\textrm{H}}(\tau)$ with error up to order $O(J^2 \tau^2)$. Here  $U_{\textrm{H}}(\tau)$ is the time evolution operator of simulators with Heisenberg interactions,
\begin{equation}
\begin{split}
U_{zz}(\tau) &\approx \left(\bigotimes_j e^{i\frac{\pi}{2} \sigma_{2j}^x} \right) \left(\bigotimes_j e^{i\frac{\pi}{2} \sigma_{2j+1}^y} \right) U_{\textrm{H}}\left(\frac{\tau}{2}\right) \\
& \quad \, \left(\bigotimes_j e^{-i\frac{\pi}{2} \sigma_{2j}^x} \right)\left(\bigotimes_j e^{-i\frac{\pi}{2} \sigma_{2j+1}^y} \right) U_{\textrm{H}}\left(\frac{\tau}{2}\right).
\end{split}
\end{equation}
Figure~\ref{fig4}(b) shows that $U_{l,l+1}^{zz}(\tau)$ can be simulated through $U_{zz}(\tau)$, that is,
\begin{equation}
\begin{split}
U_{l,l+1}^{zz}(\tau) &= \left( \bigotimes_{j'<l} e^{i\frac{\pi}{2} \sigma_{j'}^x} \right) \left( \bigotimes_{j''>l+1} e^{i\frac{\pi}{2} \sigma_{j''}^x} \right) U_{zz}
\left(\frac{\tau}{2}\right) \\
& \quad \, \left( \bigotimes_{j'<l} e^{-i\frac{\pi}{2} \sigma_{j'}^x} \right) \left( \bigotimes_{j''>l+1} e^{i\frac{\pi}{2} \sigma_{j''}^x} \right) U_{zz}\left(\frac{\tau}{2}\right)
\end{split}
\end{equation}
where $j'$ and $j''$ denote the  $j'$th and  $j''$th qubits in the simulator. Here $j'$ is even (odd) and $j''$ is (even) when $l$ is an odd (even) number. If $O(J^2 \tau^2)$ is negligibly small, $U_{l,l+1}^{xx+yy}(\tau)$ can be obtained in the same way as it is in Eq. (\ref{eq:10}).

\begin{figure}[htb] 
  \centerline{\subfigure[]{
    \includegraphics[height=3.8cm]{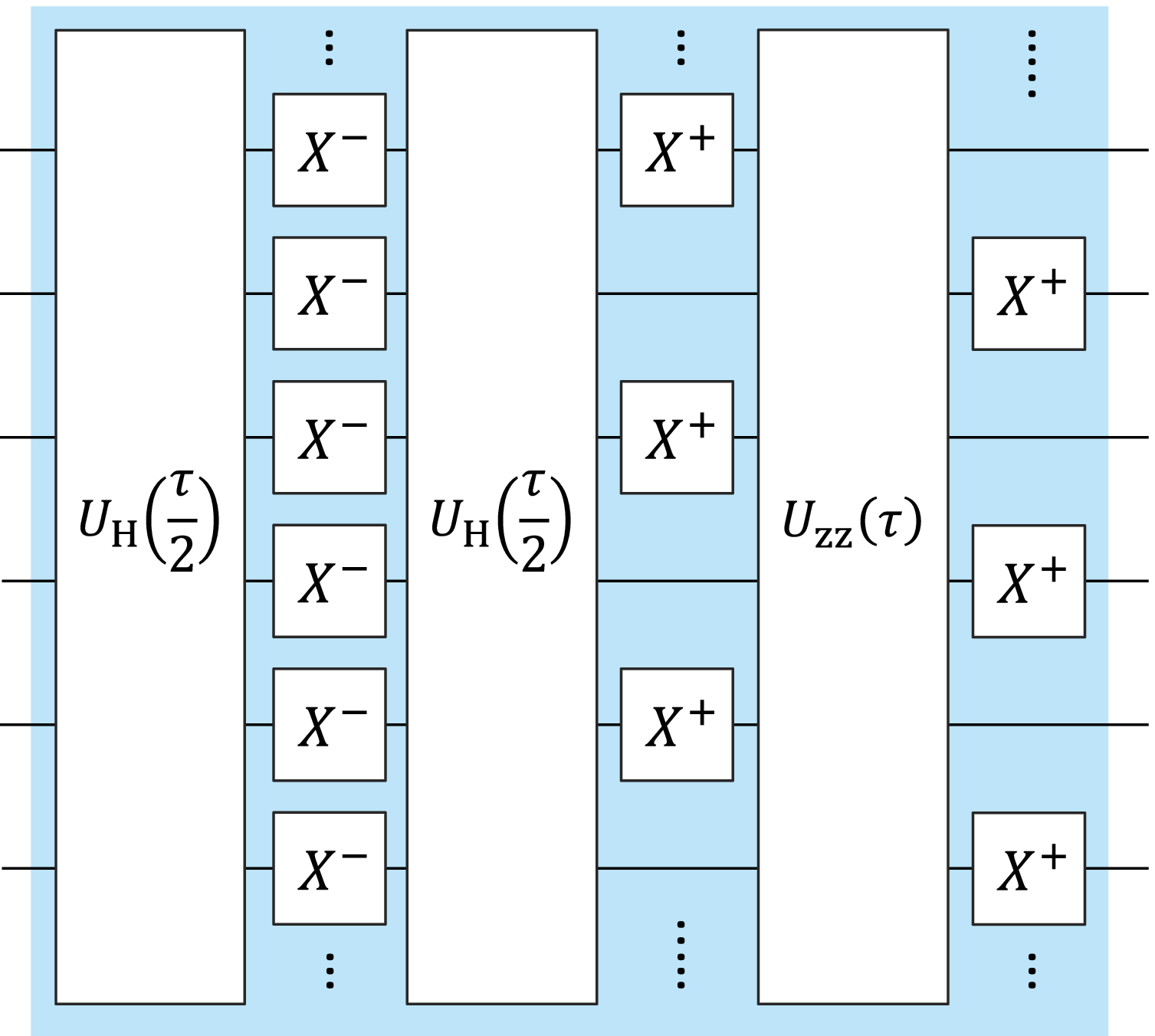}}
  \hspace{0.05cm}
  \subfigure[]{
    \includegraphics[height=3.8cm]{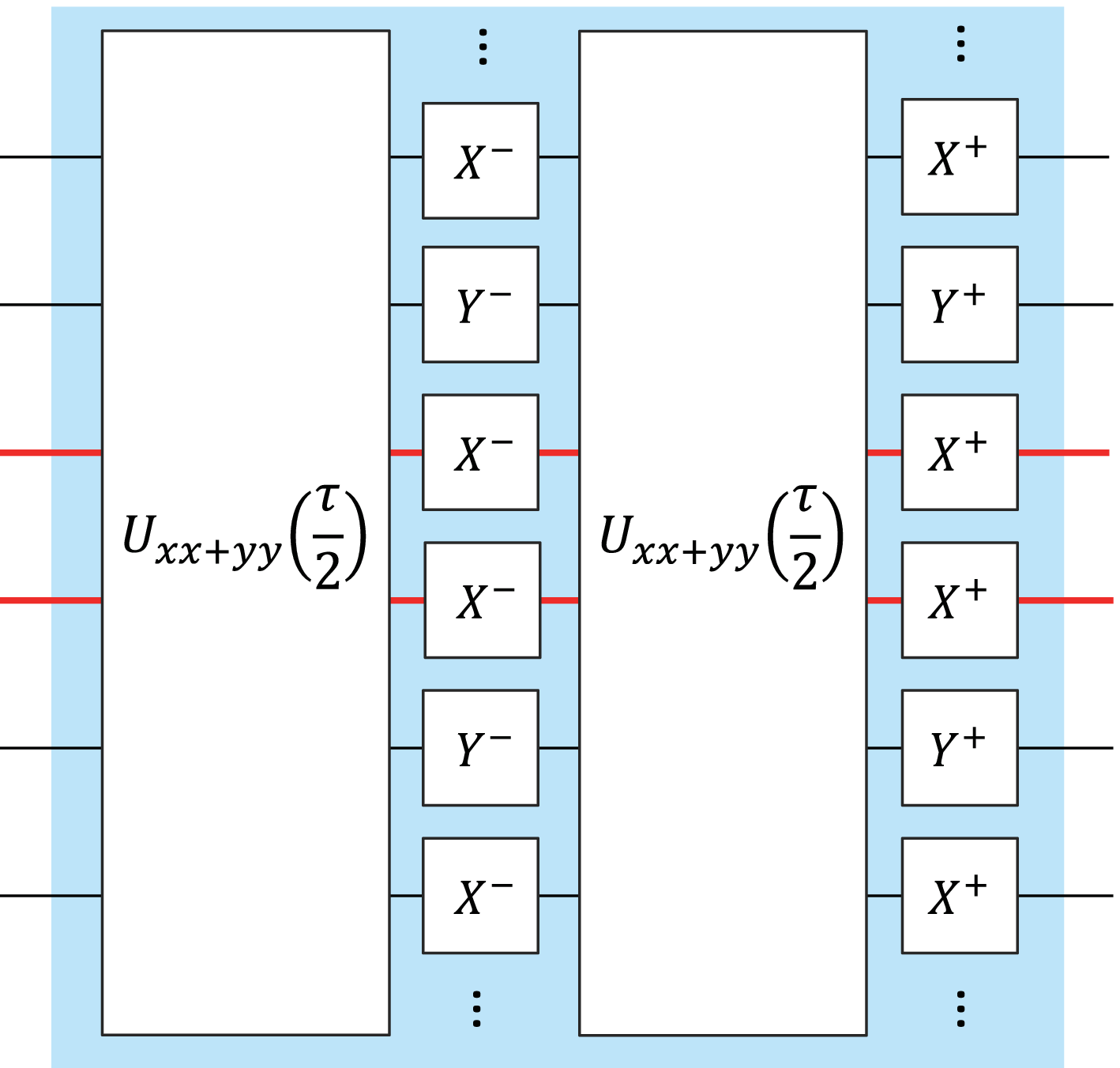}}
    }
  \caption{The quantum circuits to simulate (a) $U_{xx+yy}(\tau)$ from $U_{\textrm{H}}(\tau)$ and (b) $U_{l,l+1}^{xx+yy}(\tau)$ directly from $U_{xx+yy}(\tau)$. The red bold lines stand for the $l$th and the $(l+1)$th qubits in the system. Black dashed lines on both sides stand for the periodic extension of the logic gates to the rest of the system in the given direction. In (a), for operators $X^-$s in the left column the period is one, while for operators $X^+$s in the middle and right columns the period is two. In (b), the period is two. }\label{fig5}
\end{figure}

There is an alternative approach to simulate $U_{l,l+1}^{xx+yy}(\tau)$ through simulators with Heisenberg interactions. Consider a time evolution operator
\begin{equation}
U_{xx+yy}(\tau) = \exp\left[-i \tau \sum_{l=1}^{N-1} J_l (\sigma_l^x\sigma_{l+1}^x +\sigma_l^y\sigma_{l+1}^y )\right].
\end{equation}
Figure~\ref{fig5}(a) shows that $U_{xx+yy}(\tau)$ can be simulated through $U_{\textrm{H}}(\tau)$, that is,
\begin{equation} \label{Hei_xxyy}
\begin{split}
U_{xx+yy}(\tau) & \approx \left(\bigotimes_{j} e^{i\frac{\pi}{2} \sigma_{2j}^x} \right)U_{zz}(\tau) \left(\bigotimes_{j} e^{i\frac{\pi}{2} \sigma_{2j+ 1}^x} \right) \\
& \quad \, U_{\textrm{H}} \left(\frac{\tau}{2}\right) \left(\bigotimes_{j} e^{-i\frac{\pi}{2} \sigma_{j}^x}  \right) U_{\textrm{H}} \left(\frac{\tau}{2}\right).
\end{split}
\end{equation}
Then $U_{l,l+1}^{xx+yy}(\tau)$ can be realized from $U_{xx+yy}(\tau)$ as in Fig.~\ref{fig5}(b), which is given as
\begin{equation}
\begin{split}
U_{l,l+1}^{xx+yy}(\tau) &\approx \left(\bigotimes_{j''\leq l}e^{i\frac{\pi}{2} \sigma_{j''}^x}\right) \left(\bigotimes_{j'>l+1}e^{i\frac{\pi}{2} \sigma_{j'}^x}\right) \\
& \left(\bigotimes_{j'\leq l}e^{i\frac{\pi}{2} \sigma_{j'}^y}\right) \left(\bigotimes_{j''>l}e^{i\frac{\pi}{2} \sigma_{j''}^x}\right) U_{xx+yy}\left(\frac{\tau}{2}\right) \\
& \quad \, \left(\bigotimes_{j'\leq l}e^{-i\frac{\pi}{2} \sigma_{j'}^y}\right) \left(\bigotimes_{j''>l} e^{-i\frac{\pi}{2} \sigma_{j''}^x}\right) \\
& \left(\bigotimes_{j''\leq l}e^{-i\frac{\pi}{2} \sigma_{j''}^x}\right) \left( \bigotimes_{j'>l+1}  e^{-i\frac{\pi}{2} \sigma_{j'}^x} \right)  U_{xx+yy}\left(\frac{\tau}{2}\right).
\end{split}
\end{equation}
The approximation is valid when $O(J^2 \tau^2)$ is negligible. $j'$ is even (odd) while $j''$ is odd (even) if $l$ is odd (even), where $j''$ may equal to $l$.

Furthermore, since the Heisenberg interactions can be converted to the longitudinal Ising Hamiltonians in Eq.~(\ref{Hei_L2}), after this conversion the simulation can also be done according to the algorithm for simulators with longitudinal Ising Hamiltonians.

\subsection{Algorithm for Simulators with the XY  Nearest-Neighbor Interaction} \label{solu_XY}

\subsubsection{Simulation Algorithm for $H_{p0}$}

\begin{figure}[htb] 
  \centerline{
  \subfigure[]{
    \includegraphics[height=3cm]{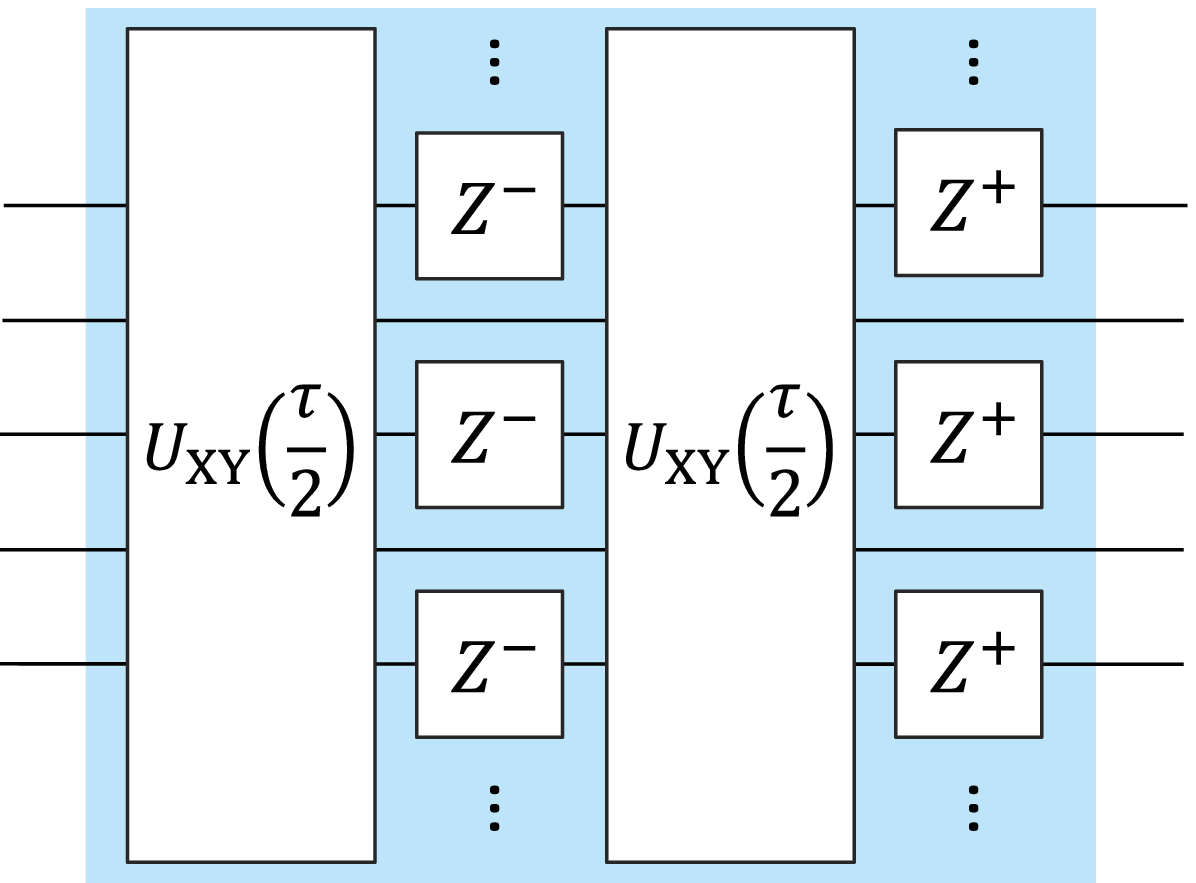}}
  \hspace{0.05cm}
  \subfigure[]{
    \includegraphics[height = 3cm]{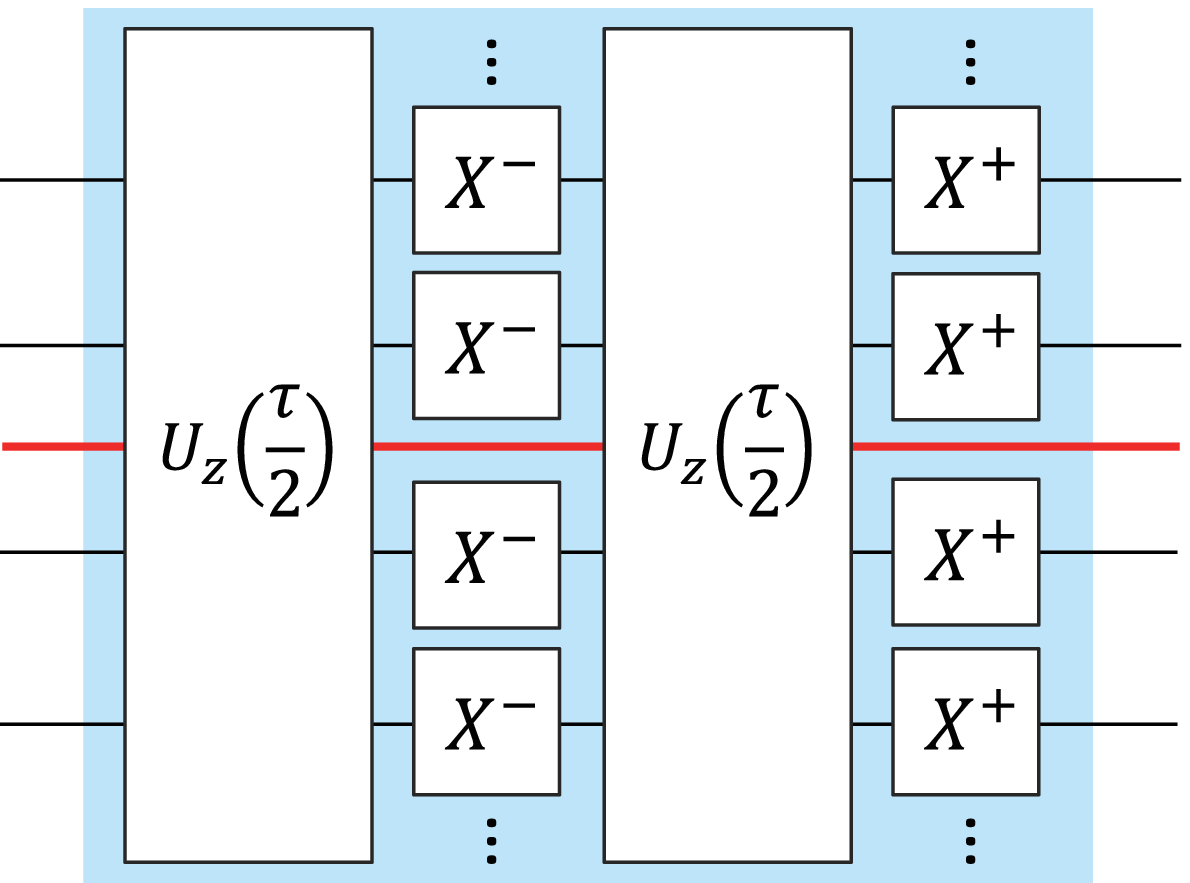}}
  }
  \caption{The quantum circuit for achieving (a) $U_{z}(\tau)$ from $U_{\textrm{XY}}(\tau)$ and (b) $U_l^z(\tau)$ from $U_{z}(\tau)$. The red bold line stands for the $l$th qubit in the system. The period of extension represented by black dots is two in (a), while is one in (b).}\label{fig6}
\end{figure}

If simulators with the XY interaction are available and the transition frequencies of all qubits in the simulators are identical or almost identical,  the time evolution operator is separable,
\begin{eqnarray}
U_{\textrm{XY}}(\tau) =e^{-i\tau H_{\textrm{XY}}} =\exp ( -i\tau H_{0})
\exp\left[-i\tau\sum_{l=1}^{N-1}H_{l}^{xy}\right].\label{eq:22}
\end{eqnarray}
Here $H_{0}$ and $H_{l}^{xy}$ are given in Eq.~(\ref{eq:4}) and Eq.~(\ref{eq:11}) respectively. Figure~\ref{fig6}(a) shows when the Heisenberg Hamiltonian in Fig.~\ref{fig3} is replaced by the XY type Hamiltonian, we can obtain $U_z(\tau) = \exp (-i \tau\sum_{l=1}^N \omega_l \sigma_l^z / 2)$, which can be expressed as
\begin{equation}\label{XY_L}
\begin{split}
U_{z}(\tau) = \left(\bigotimes_j e^{i\frac{\pi}{2}\sigma_{2j}^z}\right) U_{\textrm{XY}} \left(\frac{\tau}{2}\right) \left(\bigotimes_je^{-i\frac{\pi}{2}\sigma_{2j}^z}\right) U_{\textrm{XY}} \left(\frac{\tau}{2}\right).
\end{split}
\end{equation}
The quantum circuit for $U_{l}^{z}(\tau)$ is shown in Fig.~\ref{fig6}(b) and can be expressed as
\begin{equation} \label{XY_L2}
U_l^z(\tau) = \left(\bigotimes_{j\neq l}e^{i\frac{\pi}{2}\sigma_{j}^x}\right) U_{z} \left(\frac{\tau}{2}\right) \left(\bigotimes_{j\neq l}e^{-i\frac{\pi}{2}\sigma_{2j}^x}\right) U_{z} \left(\frac{\tau}{2}\right).
\end{equation}
The single-qubit Hamiltonian $H_{p0}$ can be simulated in the same way from $U_l^z(\tau)$ as those for longitudinal Ising and Heisenberg interaction.

\subsubsection{Simulation Algorithm for Interaction $H_{pI}$}

\begin{figure}[htb] 
  \centerline{
  \subfigure[]{
    \includegraphics[height=4.2cm]{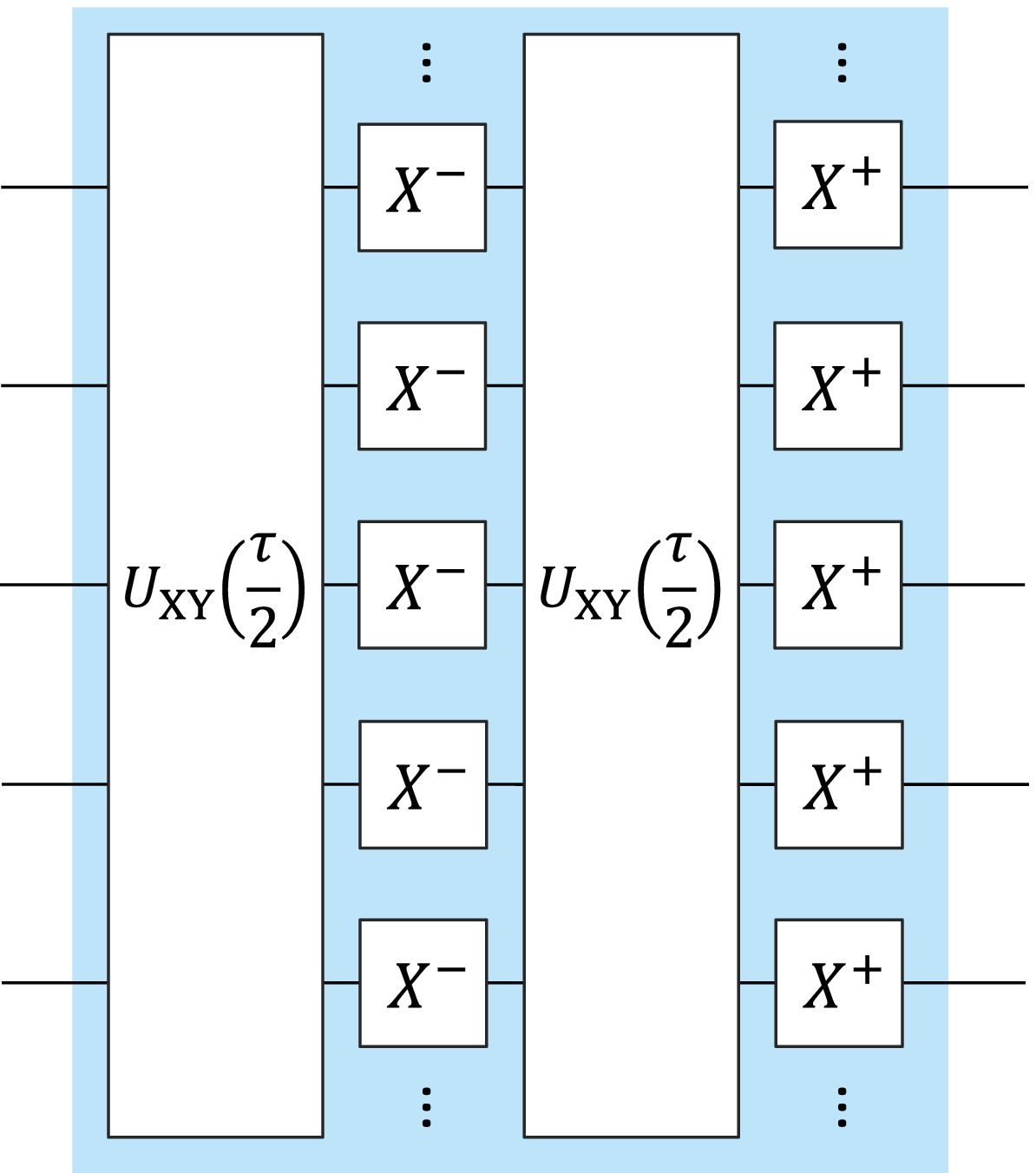}}
  \hspace{0.05cm}
  \subfigure[]{
 \includegraphics[height = 4.2cm]{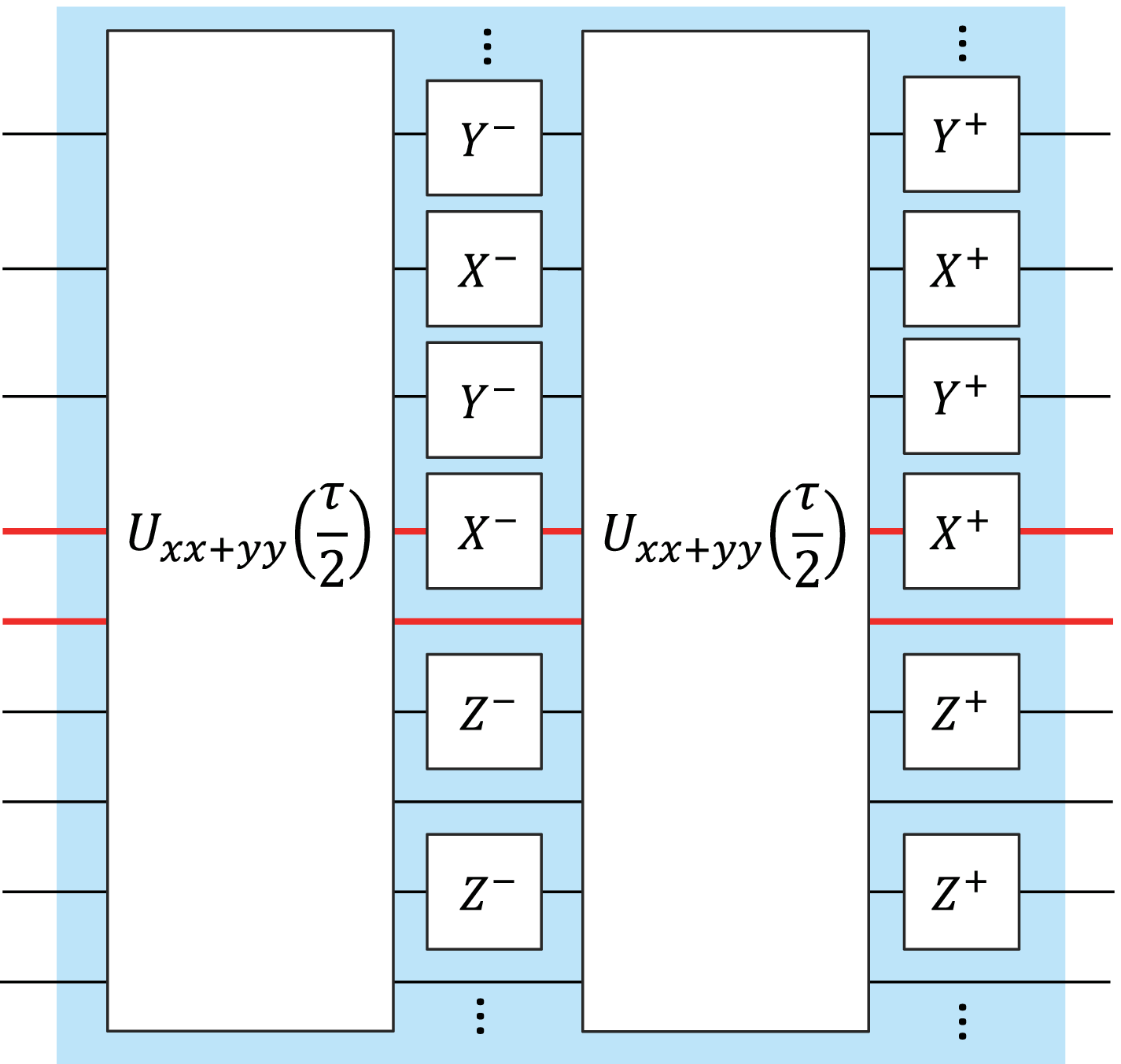}} }
  \caption{The quantum circuits for simulating (a) $U_{xx+yy}(\tau)$ from $U_{\textrm{XY}}(\tau)$ and (b) $U_{l,l+1}^{xx}(\tau)$ directly from $U_{xx+yy}(\tau)$. The red bold lines stand for the $l$th and the $(l+1)$th qubits in the system. Black dashed lines on both sides stand for the periodic extension of the logic gates. In (a) the period is one while in (b) the period is two. }\label{fig7}
\end{figure}

If we use simulators with XY interactions, then $U_{xx+yy}(\tau)$ can be simulated in the following way
\begin{equation} \label{xy_xxyy}
\begin{split}
U_{xx+yy}(\tau) = \left(\bigotimes_{j} e^{i\frac{\pi}{2} \sigma_{j}^x}\right) U_{\textrm{XY}} \left(\frac{\tau}{2}\right) \left(\bigotimes_{j}e^{-i\frac{\pi}{2} \sigma_{j}^x}\right) U_{\textrm{XY}}\left(\frac{\tau}{2}\right),
\end{split}
\end{equation}
which is shown in Fig.~\ref{fig7}(a).

Figure~\ref{fig7}(b) shows that we can acquire $U_{l,l+1}^{xx}(\tau)$ from $U_{xx+yy}(\tau)$ in the following way
\begin{equation} \label{xxyy_xxll}
\begin{split}
U_{l,l+1}^{xx}(\tau) & = \left(\bigotimes_{j''}e^{i\frac{\pi}{2} \sigma_{j''}^x} \right) \left(\bigotimes_{j'<l}e^{i\frac{\pi}{2} \sigma_{j'}^y}\right) \\
& \quad \, \left(\bigotimes_{j'>l+1}e^{i\frac{\pi}{2} \sigma_{j'}^y}\right) U_{xx+yy}\left(\frac{\tau}{2}\right)\\
& \quad \, \left(\bigotimes_{j'<l}e^{-i\frac{\pi}{2} \sigma_{j'}^y}\right) \left(\bigotimes_{j'>l+1}e^{-i\frac{\pi}{2} \sigma_{j'}^y}\right) \\
& \quad \, \left(\bigotimes_{j''}e^{-i\frac{\pi}{2} \sigma_{j''}^x} \right) U_{xx+yy}\left(\frac{\tau}{2}\right).
\end{split}
\end{equation}
Here  $j'$ is even (odd) and $j''$ is odd (even) when $l$ is an odd (even) number, where $j''$ may equal to $l$. We can then obtain $U_{l,l+1}^{zz}(\tau)$ in terms of
\begin{equation}
U_{l,l+1}^{zz}(\tau) = e^{i\frac{\pi}{4} ( \sigma_{l}^y + \sigma_{l+1}^y )} U_{l,l+1}^{xx}(\tau) e^{-i\frac{\pi}{4}  \sigma_{l}^y + \sigma_{l+1}^y )}.
\end{equation}
Then $U_{l,l+1}^{xx+yy}(\tau)$ can be simulated using $U_{l,l+1}^{zz}(\tau)$ according to Eq. (\ref{eq:10}).
We note that $U_{l,l+1}^{xx+yy}(\tau)$ can also be directly given through the process,
\begin{eqnarray}
U_{l,l+1}^{xx+yy}(\tau) &=& U_{l,l+1}^{xx}(\tau) e^{i\frac{\pi}{4} ( \sigma_{l}^z + \sigma_{l+1}^z )} U_{l,l+1}^{xx}(\tau) e^{-i\frac{\pi}{4} ( \sigma_{l}^z + \sigma_{l+1}^z )}\nonumber\\
& + & O(J^2 \tau^2),
\end{eqnarray}
Then the interaction Hamiltonian $H_{pI}$ can be finally simulated in the same way as it was shown in Ref.~\cite{Wu02}.

\subsection{Algorithm for Simulators with Transverse Ising Nearest-Neighbor Interaction}

\subsubsection{Simulation Algorithm for Individual $H_{p0}$}

\begin{figure}[htb] 
  \centerline{
  \subfigure[]{
    \includegraphics[height=2.65cm]{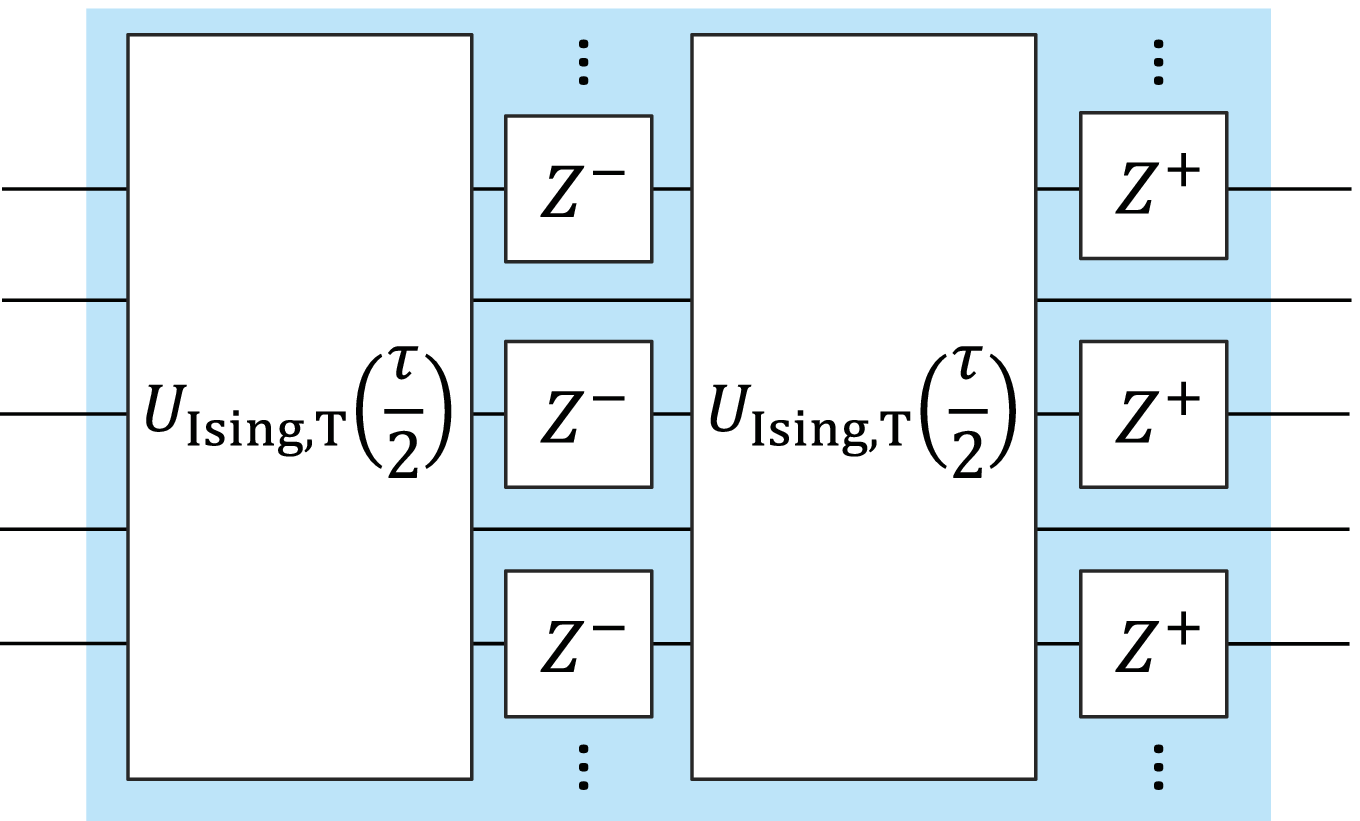}
    }
  \hspace{0.05cm}
  \subfigure[]{
    \includegraphics[height=2.65cm]{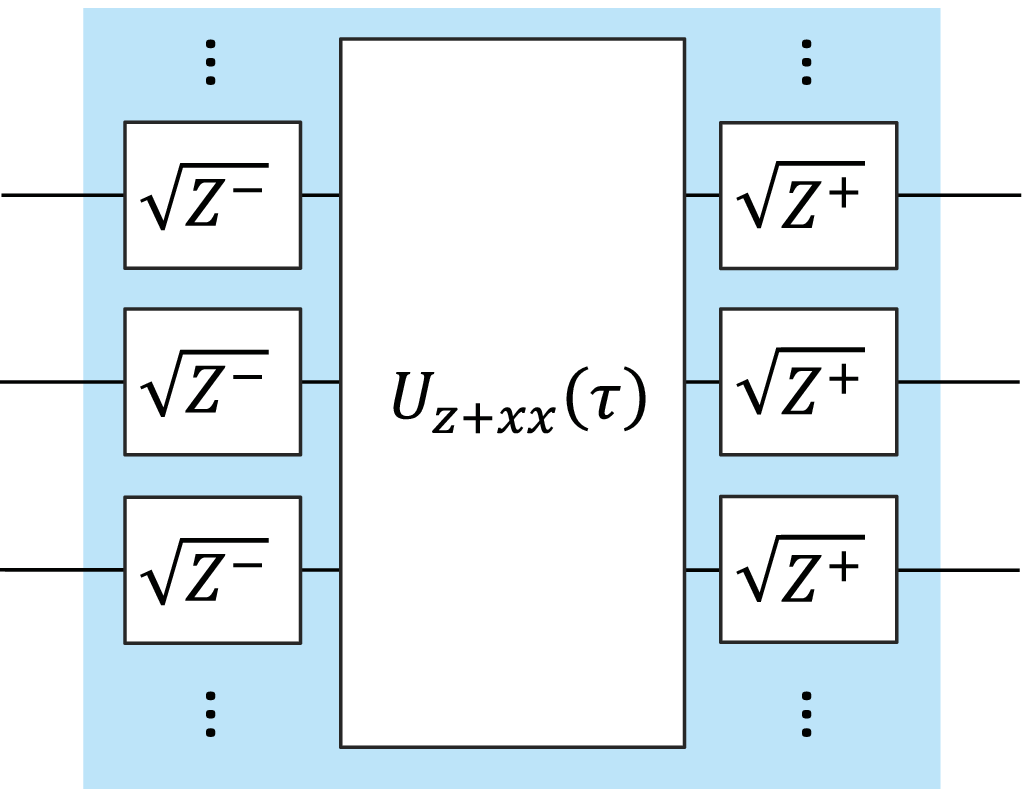}
    }}
  \caption{The quantum circuit for simulating (a) $U_z(\tau)$ from $U_{\textrm{Ising,T}}$ and (b) $U_{z+yy}(\tau; \omega_1,\dots, \omega_N)$ from $U_{z+xx}(\tau; \omega_1,\dots, \omega_N)$. Black dots on both sides stand for the extension of the logic gates with period one. $\sqrt{Z^\pm}$ stand for external gates $e ^ {\pm i\frac{\pi}{4} \sigma^z}$.}\label{fig8}
\end{figure}

In contrast to the above cases, $U_{\textrm{Ising,T}}(\tau) = e^{-i\tau H_{\textrm{Ising,T}}}$ for the Hamiltonian with transverse Ising interactions  is not exactly separable. Even so, as shown in Fig.~\ref{fig8}(a), $U_z(\tau)$ can be extracted by the same quantum circuit as it was in Fig.~\ref{fig6}(a) and Fig.~\ref{fig6}(b), if error up to order $O(\omega J \tau^2 )$ is tolerable. Here $\omega$ is the typical value of qubit frequency. However, we will show in Sec.~\ref{numerical} that this approximation due to Trotter's formula leads to the fluctuation in fidelity.

\subsubsection{Simulation Algorithm for Interaction $H_{pI}$}

Let us first consider a time evolution operator
\begin{equation}
U_{z+xx}(\tau)  = \exp \left[ -i \tau\left(\sum_{l=1}^N \frac{\omega_l }{2}\sigma_l^z  + \sum_{l=1}^{N-1} J_l \sigma_l^x \sigma_{l+1}^x \right) \right],
\end{equation}
which is exactly the same as $U_{\textrm{Ising,T}}(\tau)$ with tunable parameters $\omega_l$ $(l = 1,\dots, N)$. Similarly, we also consider
\begin{equation}
U_{z+yy}(\tau)= \exp \left[ -i \tau \left( \sum_{l=1}^N \frac{\omega_l}{2} \sigma_l^z + \sum_{l=1}^{N-1} J_l \sigma_l^y \sigma_{l+1}^y \right) \right].
\end{equation}
The operation $U_{z+yy}(\tau)$ can be acquired from $U_{z+xx}(\tau)$ by
\begin{equation}
\begin{split}
U_{z+yy}(\tau) = \left(\bigotimes_{j=1}^N e^{i\frac{\pi}{4}\sigma_{j}^z}\right) U_{z+xx}(\tau) \left(\bigotimes_{j=1}^N e^{-i\frac{\pi}{4}\sigma_{j}^z}\right),
\end{split}
\end{equation}
which is graphically shown in Fig.~\ref{fig8}(b). If all qubit resonance frequencies in $U_{z+xx}(\tau)$ are set to be $\omega$ and those in $U_{z+yy}(\tau)$ to be $-\omega$,  we have
\begin{equation}
U_{xx+yy}(\tau) \approx U_{z+xx}(\tau) U_{z+yy}(\tau)
\end{equation}
up to order $O(J^2 \tau^2)$. 
Here, the negative qubit resonance frequencies mean that inverse external fields are used to flip the sign of $\sum_{l=1}^N \omega_l \sigma_l^z / 2$. Thus, the simulation using the transverse Ising Hamiltonian is converted to the the cases of Heisenberg or XY Hamiltonian as studied before.

\section{Numerical Study of Fidelity} \label{numerical}

We now come to study numerically the error introduced by Trotter's formula, which are typically in the order $O(J^2 t^2)$ or $O(\omega J t^2)$. Our focus will be the fidelity of the simulated Hamiltonian using the simulation algorithms.

\begin{figure}
  \includegraphics[bb=108 212 470 545, width=4cm, clip]{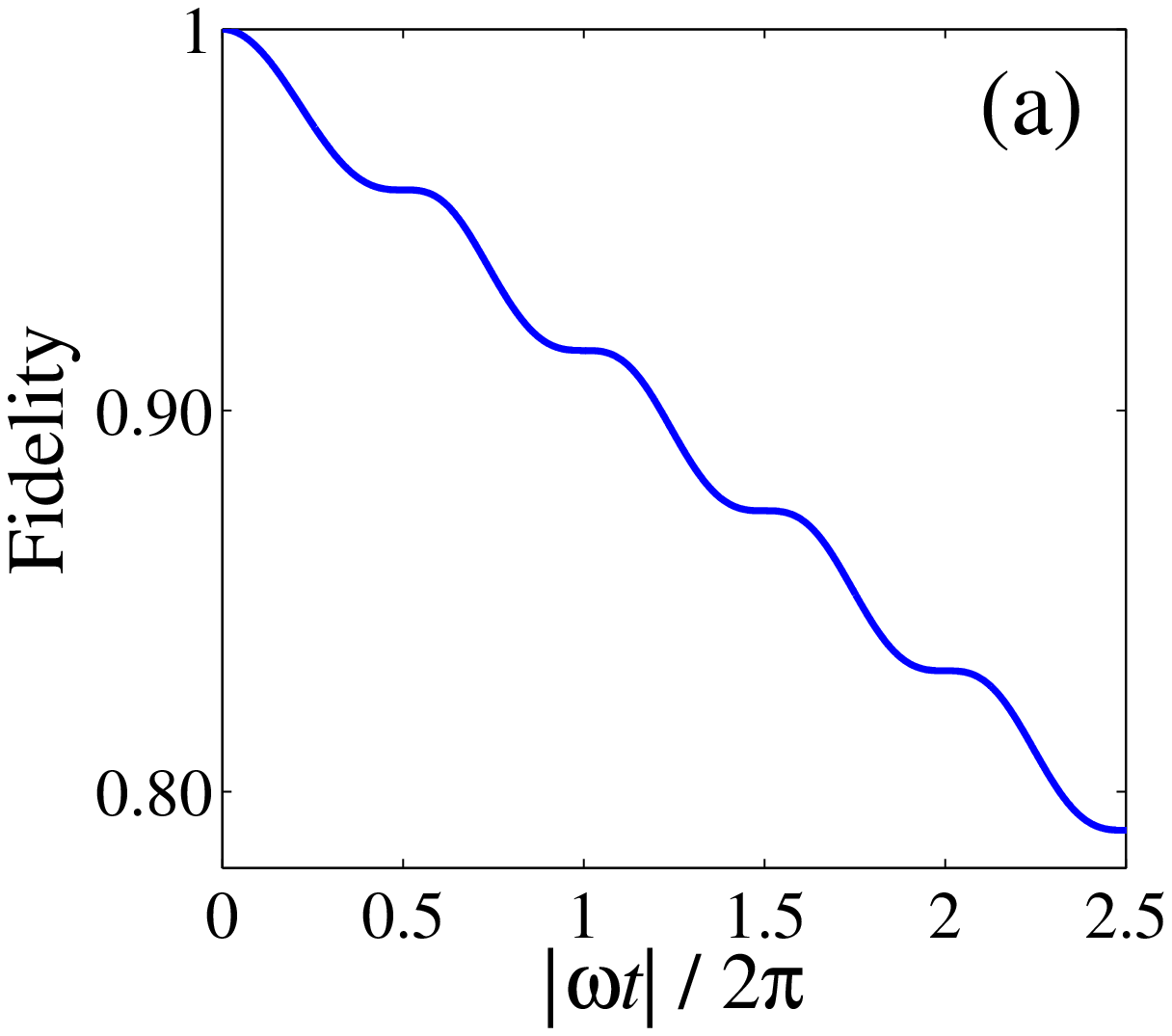}
  \includegraphics[bb=108 212 470 545, width=4cm, clip]{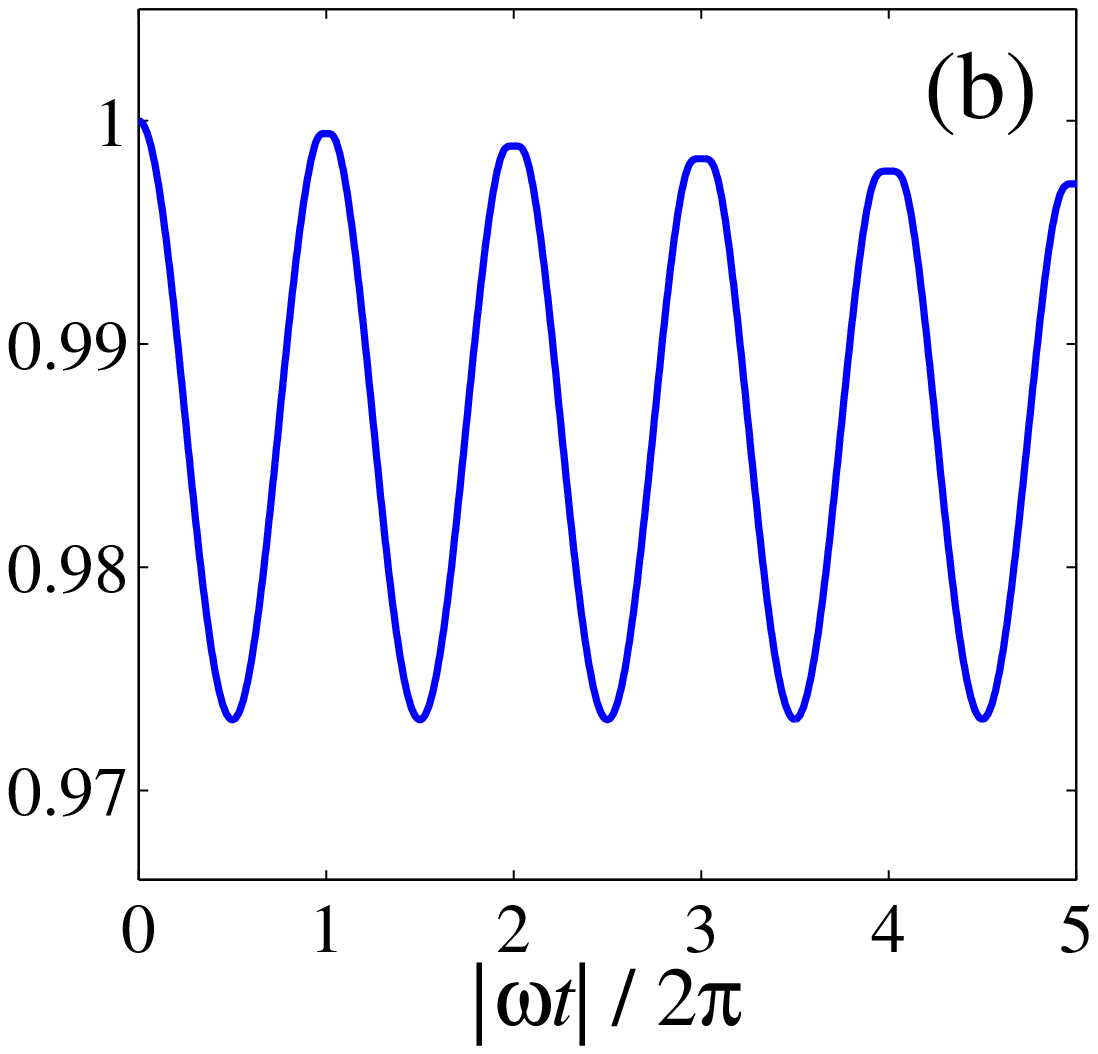}
  \caption{Digital fidelity of (a) $e^{-i\tau H_{\textrm{Ising,T}}} \approx \exp(-i\tau \sum_{l=1}^N \omega_l \sigma_l^z/ 2)$ $\exp(\sum_{l=1}^{N-1}J_l (\sigma_l^x\sigma_{l+1}^x +\sigma_l^y\sigma_{l+1}^y+\sigma_l^z\sigma_{l+1}^z)$ and (b) process shown in Fig.~\ref{fig8}(a) for a 4-qubit quantum simulator. $\varepsilon_l = 2\times10^{12}$Hz, $\omega_l = 5\times10^9$Hz $(l = 1, \dots, N)$, $V_l = -2\times10^8$Hz and $J_l = 3\times10^7$Hz $(l = 1, \dots, N-1)$. The fluctuations on the curves originate from the trigonometric terms in Eq.~(\ref{fluct}) with period $\pi/\omega$ in (a) and $2\pi / \omega$ in (b) respectively.}
\label{fig9}
\end{figure}

\begin{figure}
  \includegraphics[bb=108 212 470 545, width=4cm, clip]{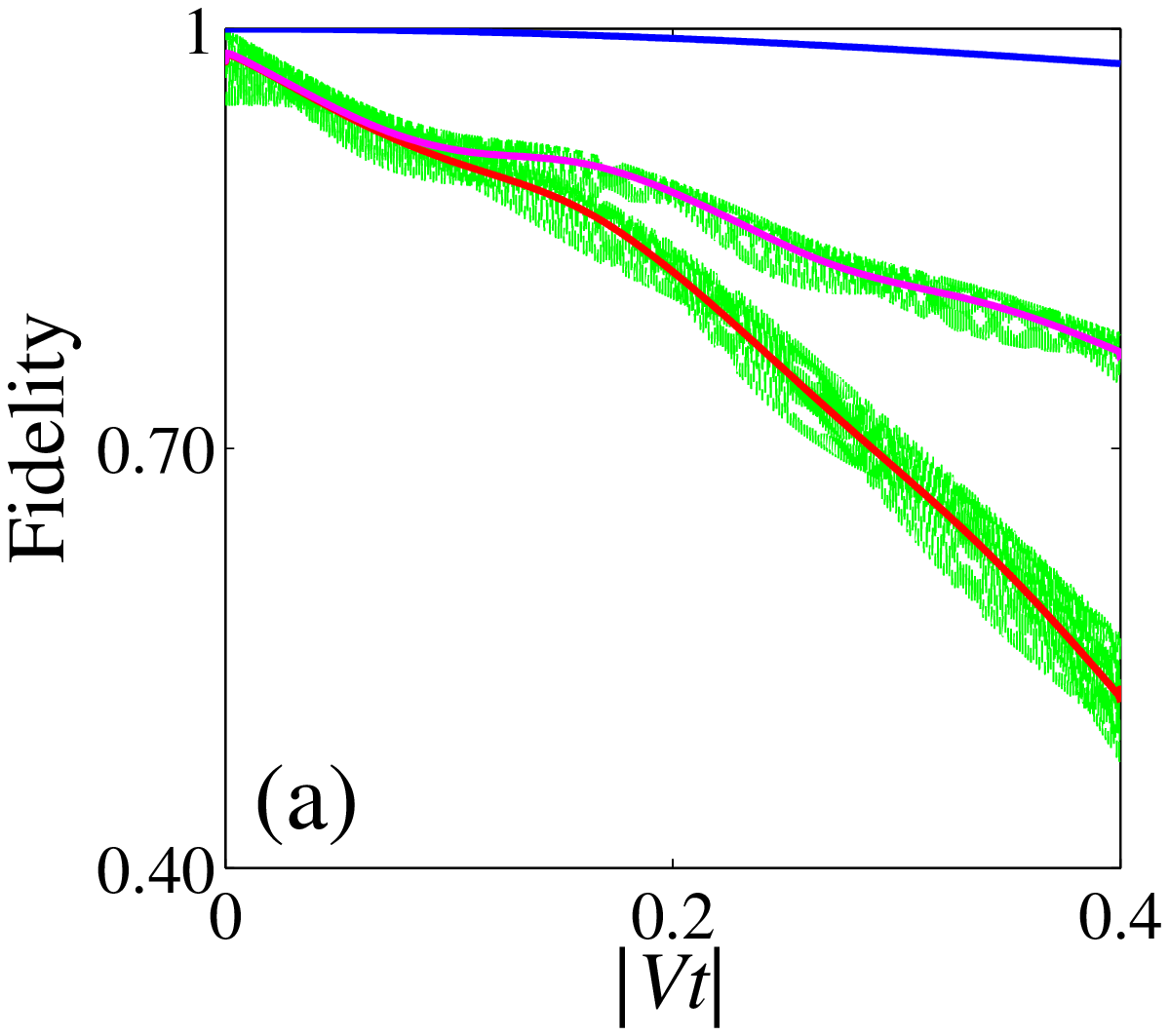}
  \includegraphics[bb=108 212 470 545, width=4cm, clip]{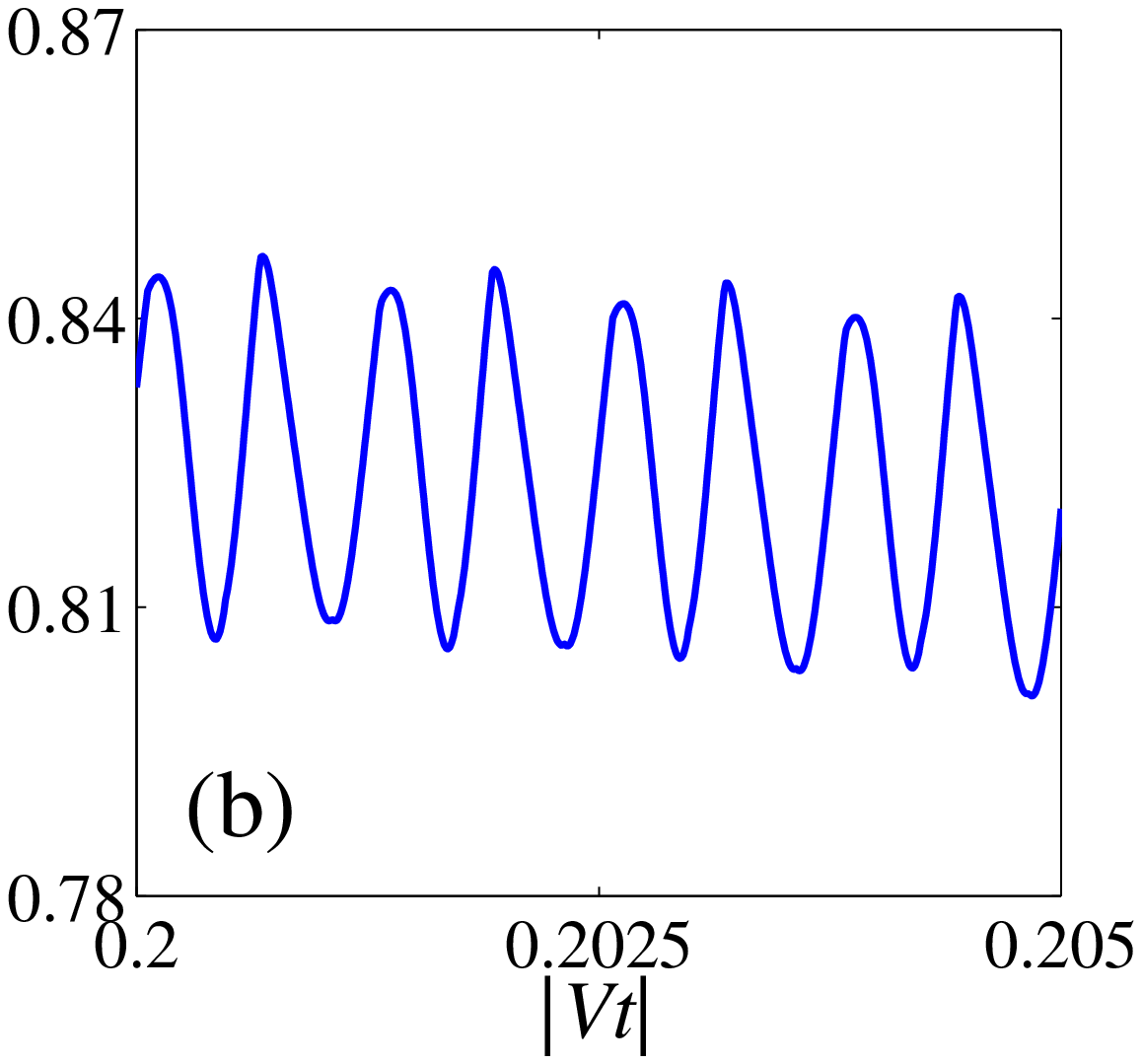}
  \includegraphics[bb=108 212 470 545, width=4cm, clip]{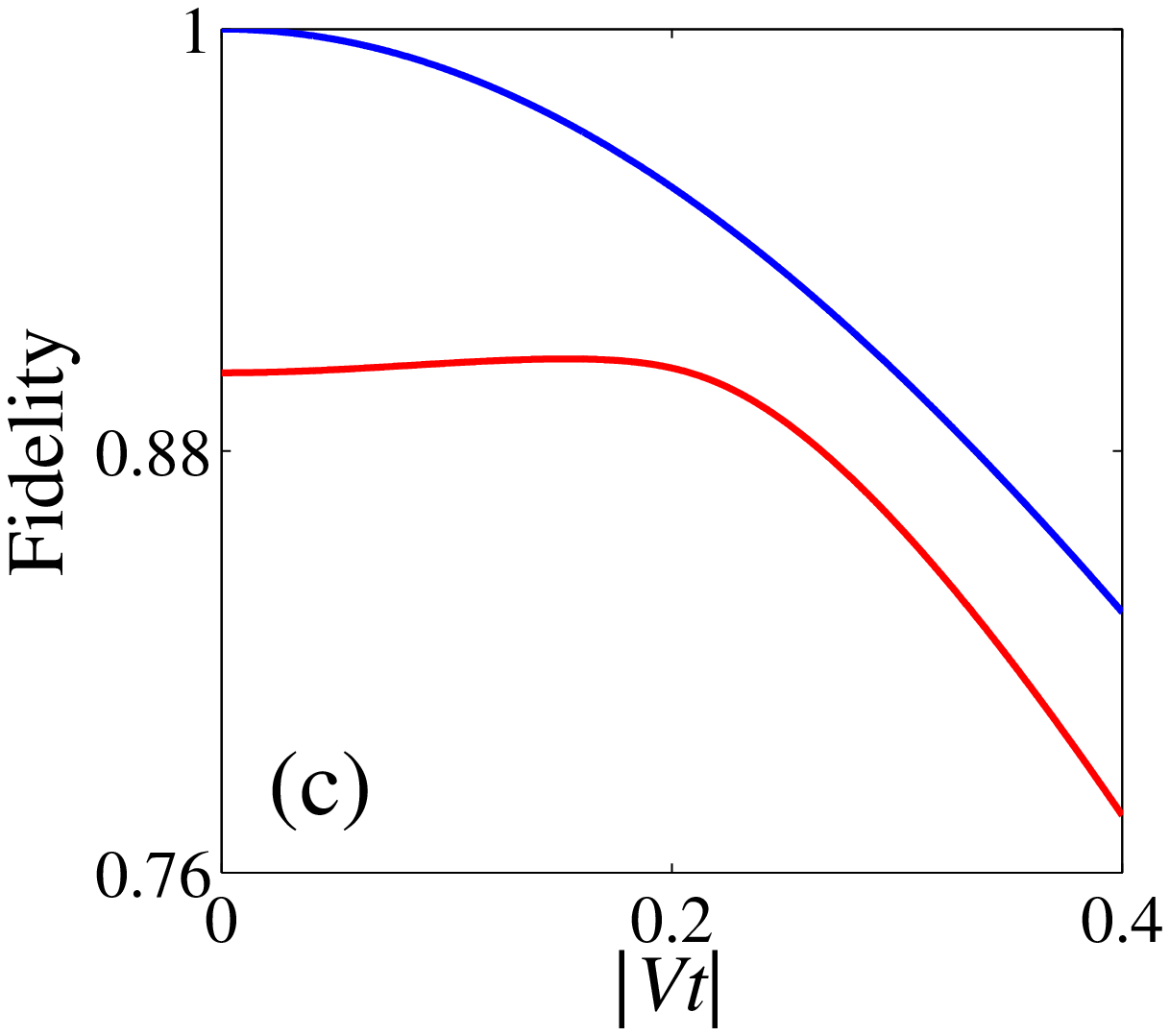}
  \includegraphics[bb=108 212 470 545, width=4cm, clip]{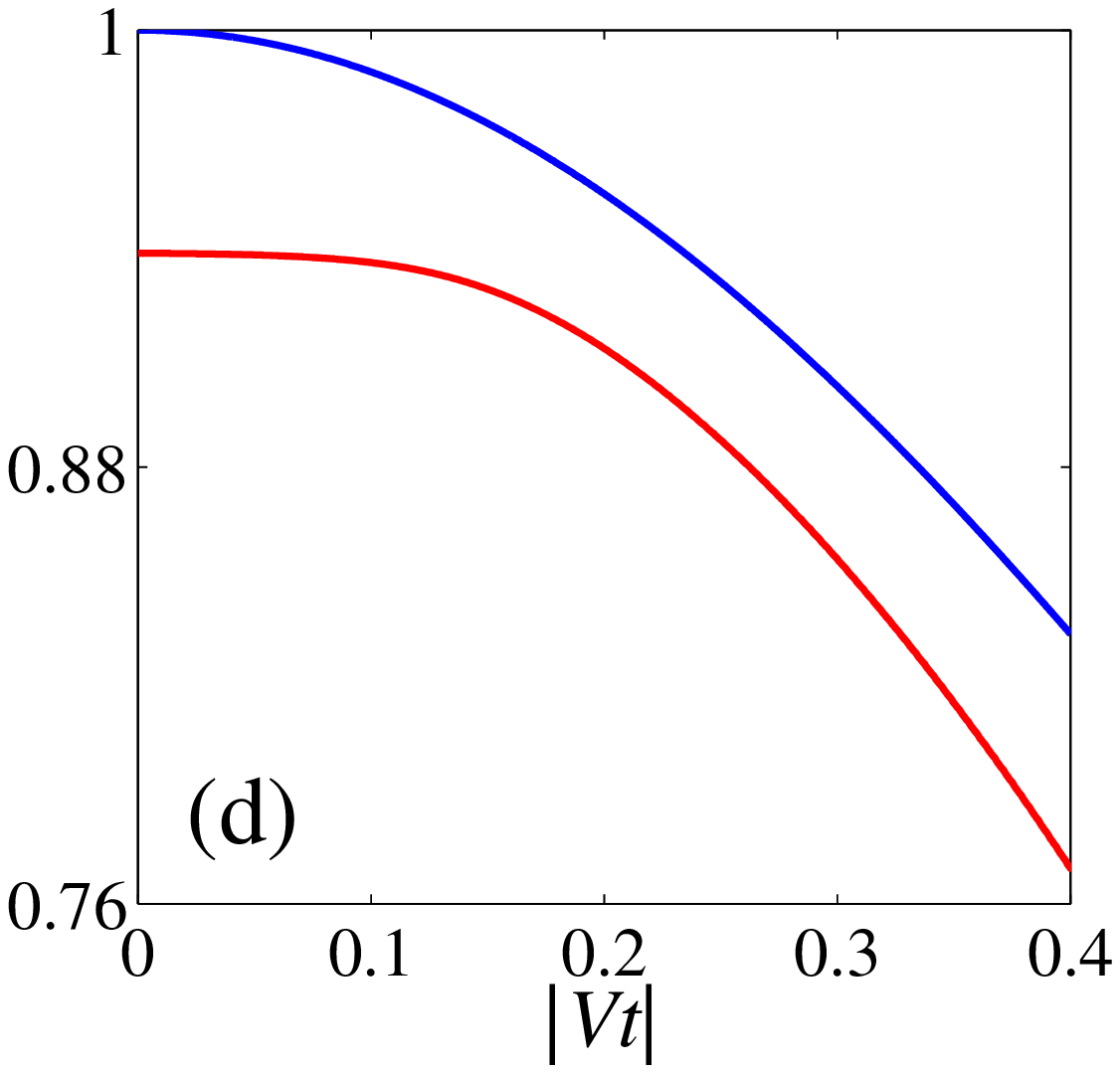}
  \caption{Increase of simulation fidelity contributed by tunable parameters of 4-qubit simulators containing (a)(b) transverse Ising type, (c) XY type and (d) Heisenberg type nearest-neighbor interaction. In all four subfigures, $\varepsilon_l = 2\times10^{12}$Hz, $\omega_l = 5\times10^9$Hz $(l = 1, \dots, N)$, $V_l = -2\times10^8$Hz. In (a)(b), $J_l = 3\times10^7$Hz $(l = 1, \dots, N-1)$ and the whole simulation task is divided into $M=20$ intervals, while in (c)(d) $J_l = 3\times10^6$Hz $(l = 1, \dots, N-1)$ and the whole simulation task is divided into $M=10$ intervals. Furthermore, the subprocess in (c) and (d) for simulators with constant parameters to simulate $e^{\pm i\tau (\sigma_l^x \sigma_{l+1}^x + \sigma_l^y \sigma_{l+1}^y) \pi / 4}$ $(l = 1, \dots, N-1)$ is divided into 200 intervals in order to reduce the error introduced by short-time approximation. In (a), red solid curve stands for the average fidelity when all the parameters of the simulator are constant. Magenta solid curve stand for the average fidelity when only $\omega_l$ $(l = 1, \dots, N)$ are tunable. Green dot curves stand for the high frequency fluctuation of simulation fidelity originated from Eq. (\ref{fluct}) associated with the above two cases. Blue solid curve gives the fidelity when all the parameters of the simulator can be turned on and off during the simulation process. The detailed shape of the fidelity from a simulator with constant parameters over a short interval is shown in (b). In both (c) and (d), red solid curve stands for the average fidelity when all the parameters of the simulator are constant while blue solid curve gives the fidelity when all the parameters of the simulator can be turned on and off.}
  \label{fig10}
\end{figure}

In Sec.~\ref{algo1}, we assume for generality that all of the algorithms are applicable to simulators with constant coupling strengths and frequencies. However, if simulators with tunable parameters are available, we can not only simplify the simulation process but also reduce digital errors significantly.
For instance, $e^{-i\tau H_{\textrm{Ising,T}}} =\exp(-i\tau \sum_{l=1}^N \omega_l \sigma_l^z/ 2)$ $\exp(\sum_{l=1}^{N-1}J_l (\sigma_l^x\sigma_{l+1}^x +\sigma_l^y\sigma_{l+1}^y+\sigma_l^z\sigma_{l+1}^z) $ does not hold exactly in general. Detailed calculation indicates
\begin{equation}\label{fluct}
\begin{split}
&\;\;\;\;\exp\left(-i\tau\sum_{l=1}^{N} \frac{\omega}{2} \sigma_l^z\right) \exp\left(-i\tau\sum_{l=1}^{N-1} J \sigma_l^x \sigma_{l+1}^x\right) \\
&=\exp\left\{ -i\tau \left( \sum_{l=1}^{N} \frac{\omega}{2} \sigma_l^z + \sum_{l=1}^{N-1} J \sigma_l^x \sigma_{l+1}^x \right) \right.\\
&\quad\;\; \left.+ \frac{i\tau }{2} \sum_{l=1}^{N-1} J \left(\sigma_l^x \sigma_{l+1}^x - \sigma_l^y \sigma_{l+1}^y \right)\right.\\
&\quad\;\; \left. -\frac{i}{4\omega} \sum_{l=1}^{N-1} J \left(\sigma_l^x \sigma_{l+1}^x - \sigma_l^y \sigma_{l+1}^y \right) \sin(2\omega\tau) \right.\\
&\quad\;\; \left.+ \frac{i}{4\omega} \sum_{l=1}^{N-1} J (\sigma_l^y \sigma_{l+1}^x + \sigma_l^x \sigma_{l+1}^y) \left[\cos(2\omega\tau) - 1\right]\right\}
\end{split}
\end{equation}
where $i\tau \sum_{l=1}^{N-1} J (\sigma_l^x \sigma_{l+1}^x - \sigma_l^y \sigma_{l+1}^y )/2$ is the error that will accumulate when simulation processes. These terms containing $\sin(2\omega\tau)$ and $\cos(2\omega\tau)$ are the origins of the fluctuations of period $\pi/\omega$ on the fidelity curve, as shown in Fig.~\ref{fig9}(a). Numerical simulation shows that the fidelity of the circuit in Fig.~\ref{fig8}(a) is plotted in Fig.~\ref{fig9}(b). The period of fluctuation becomes $2\pi / \omega$ because $U_{\textrm{Ising,T}}(\tau/2)$ instead of $U_{\textrm{Ising,T}}(\tau)$ is involved. However, if the interaction in $H_{\textrm{Ising,T}}$ can be turned off when $H_{p0}$ is simulated, those fluctuations can be avoided as shown in Fig.~\ref{fig10}(a).

The fidelity of a simulation algorithm increases when tunable parameters of the simulator increase, as exemplified in Fig.~\ref{fig10}. The effect of tunable parameters is significant except for simulator with longitudinal Ising Hamiltonian, in which all terms commute with each other.  

\begin{figure}
  \includegraphics[bb=108 212 470 545, width=4cm, clip]{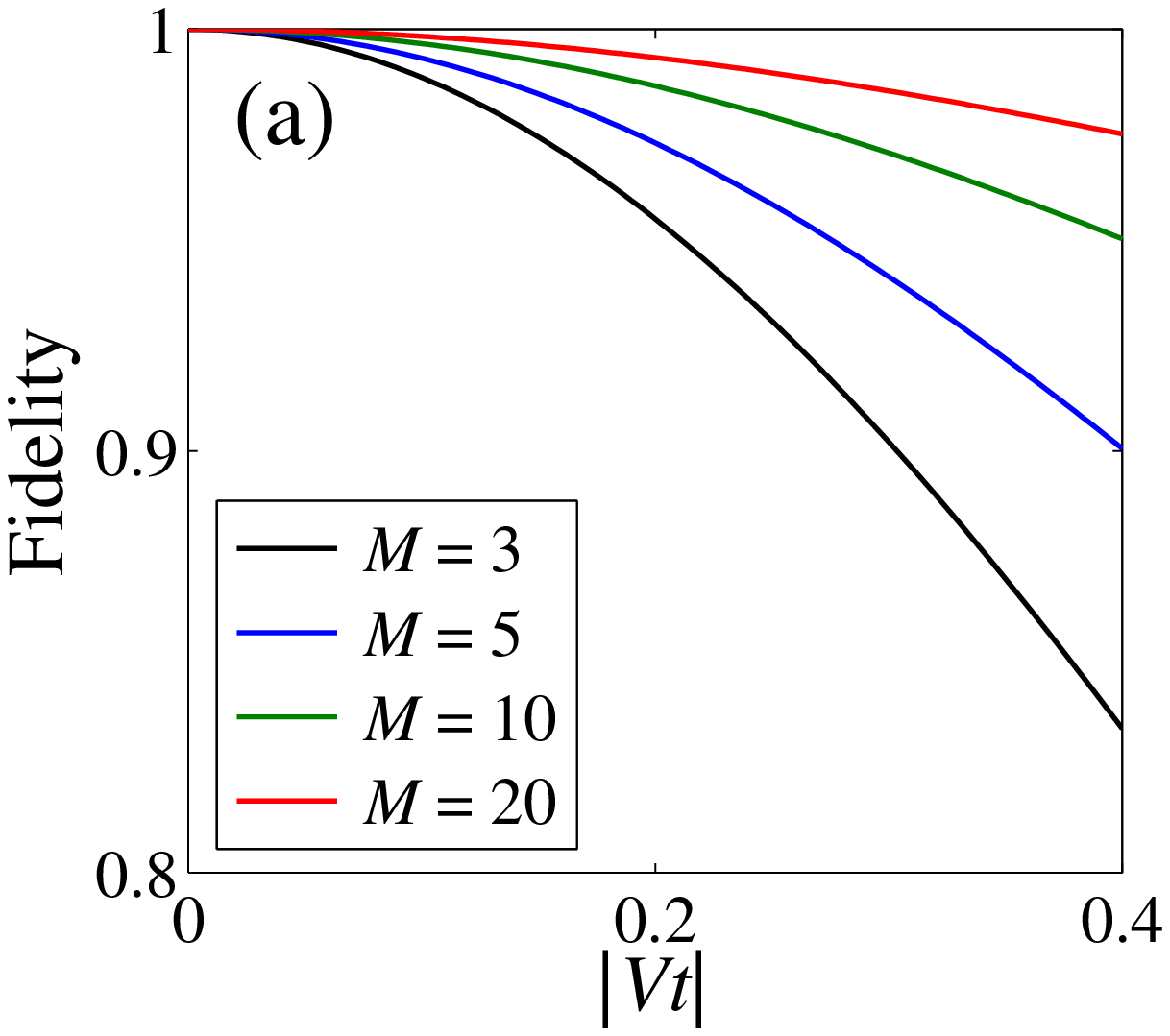}
  \includegraphics[bb=108 212 470 545, width=4cm, clip]{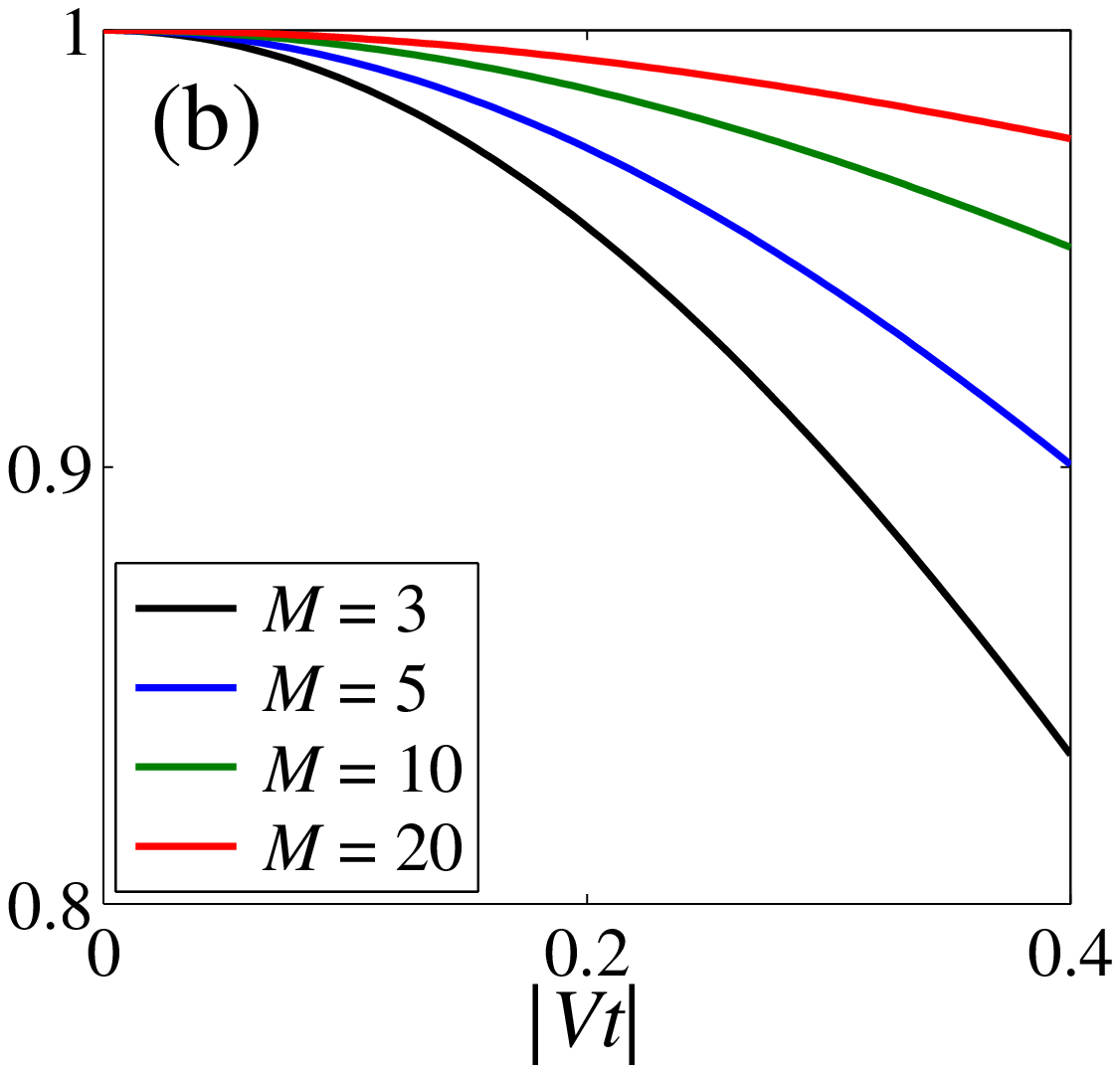}
  \includegraphics[bb=108 212 470 545, width=4cm, clip]{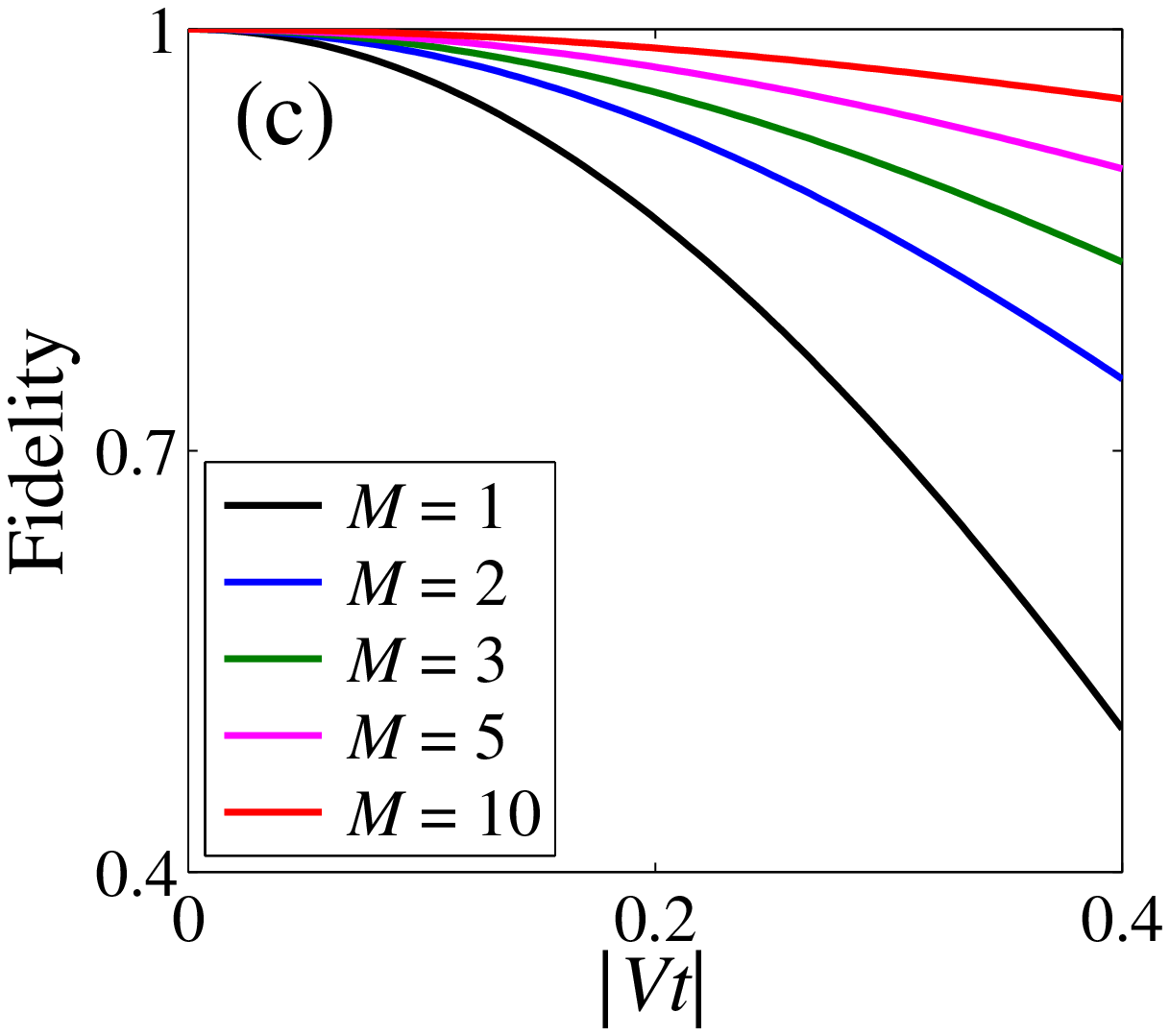}
  \includegraphics[bb=108 212 470 545, width=4cm, clip]{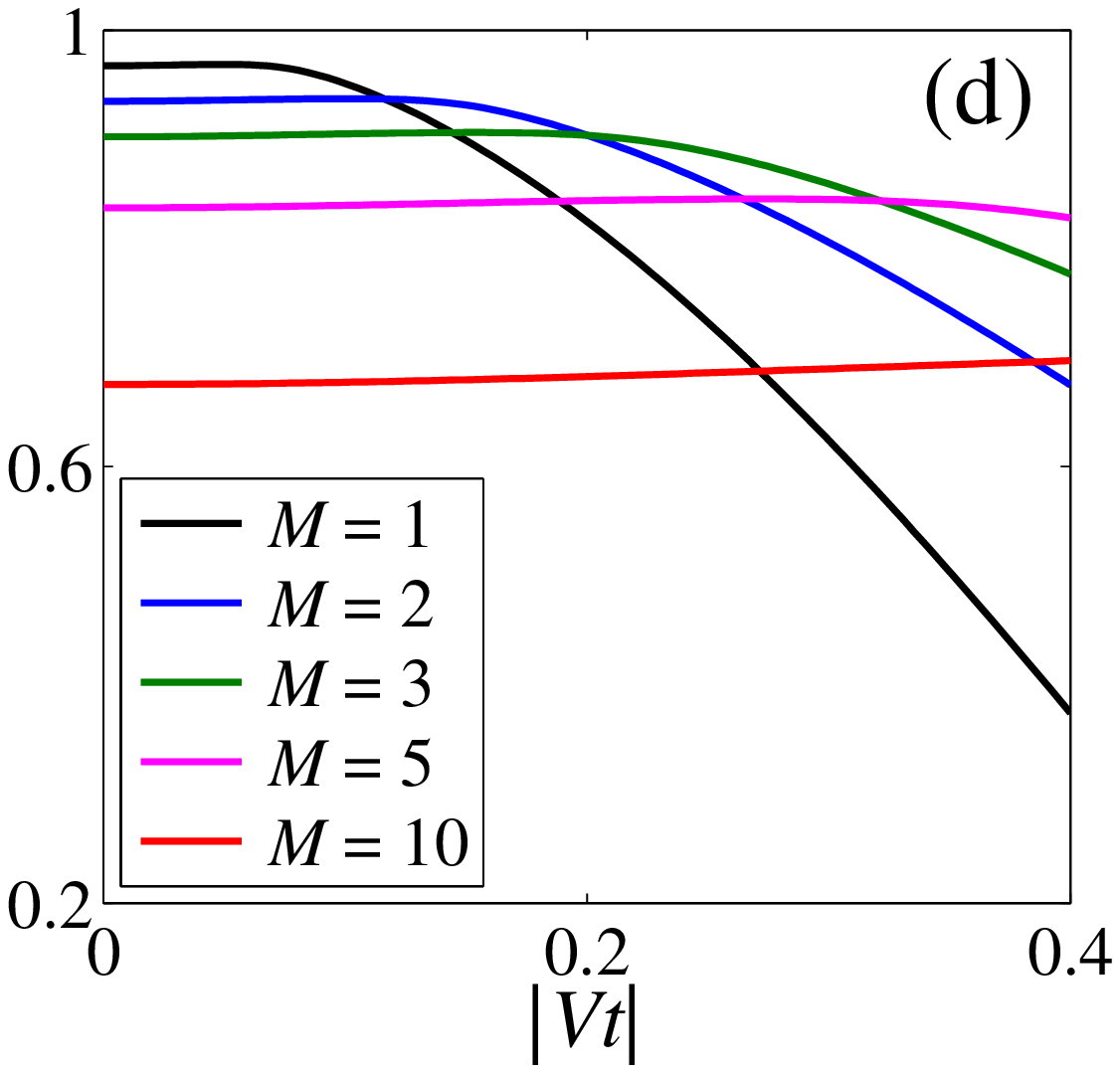}
  \includegraphics[bb=108 212 470 545, width=4cm, clip]{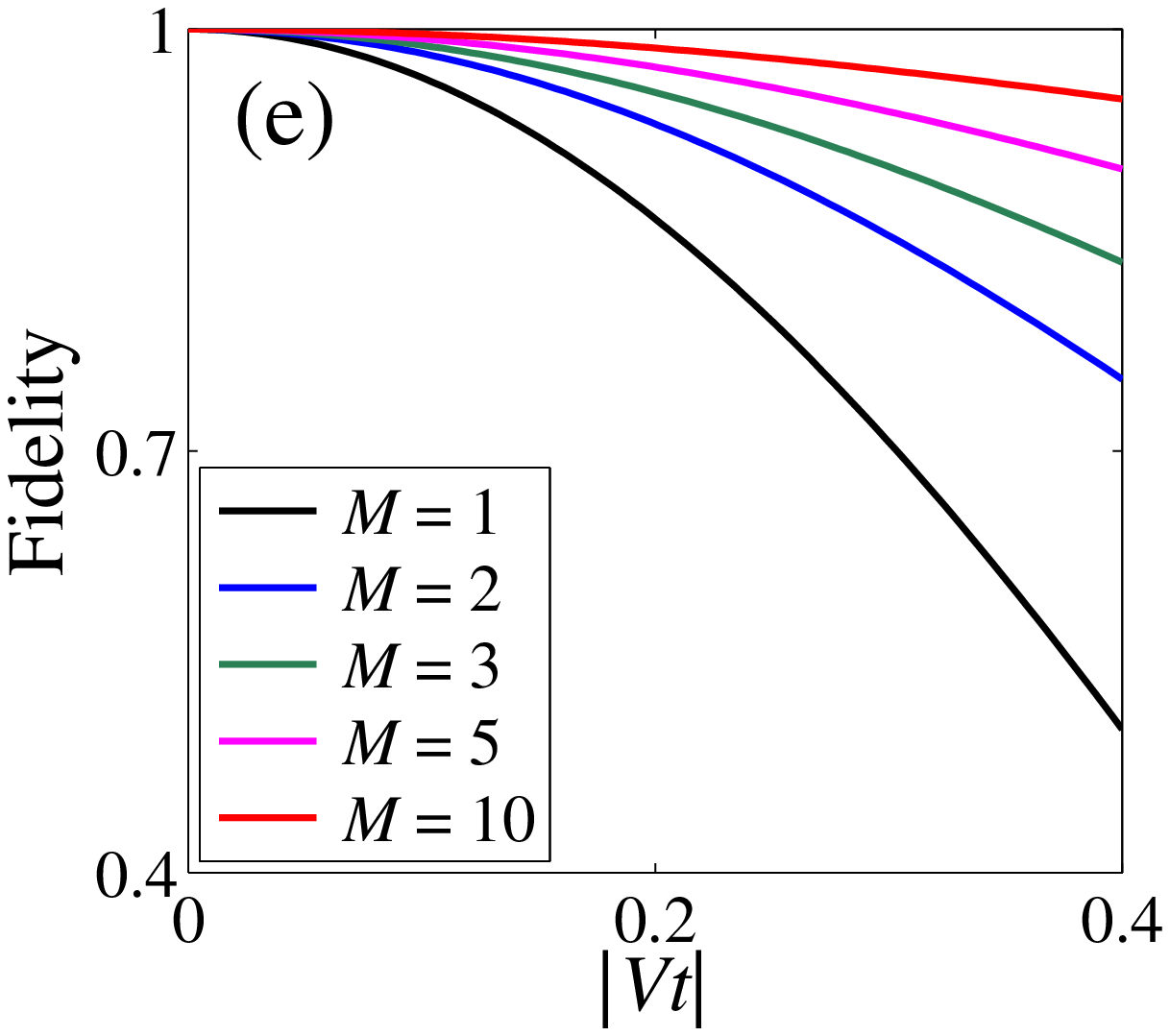}
  \includegraphics[bb=108 212 470 545, width=4cm, clip]{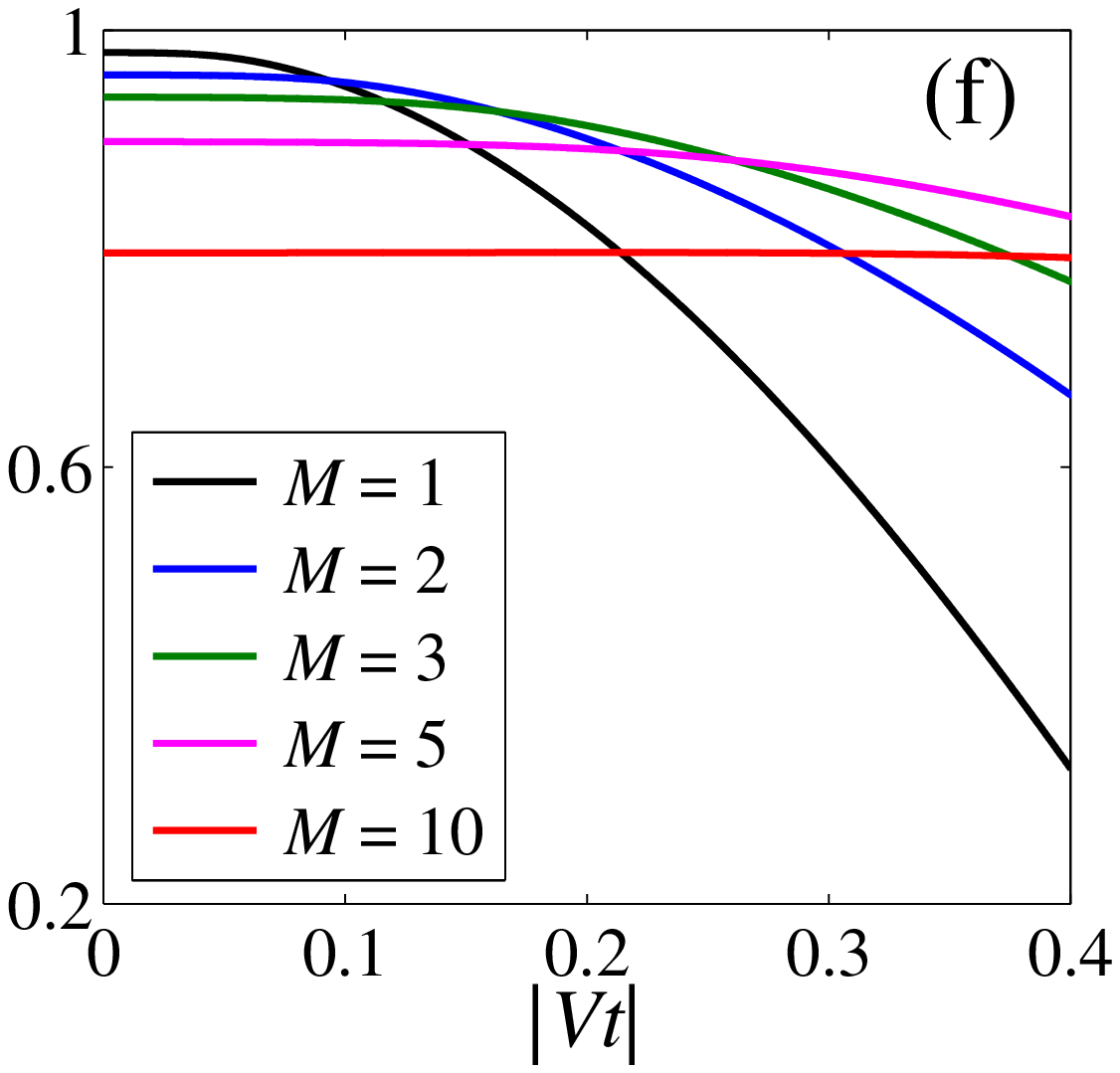}
  \caption{Effect of changing the number of intervals $M$ that the whole simulation process is divided into on the fidelity for 4-qubit simulators containing (a) longitudinal Ising type, (b) transverse Ising type, (c)(d) XY type and (e)(f) Heisenberg type nearest-neighbor interaction. In (c) and (e), all parameters of the simulators can be turned off and on during the process, while in (d) and (f) they are constant. In all six subfigures, $\varepsilon_l = 2\times10^{12}$Hz, $\omega_l = 5\times10^9$Hz $(l = 1, \dots, N)$, $V_l = -2\times10^8$Hz, while in (a)(b) $J_l = 3\times10^7$Hz and in (c)(d)(e)(f) $J_l = 1\times10^6$Hz. The subprocess in (d) and (f) for simulators with constant parameters to simulate $e^{\pm i (\sigma_l^x \sigma_{l+1}^x + \sigma_l^y \sigma_{l+1}^y) \pi / 4}$  $(l = 1, \dots, N-1)$ is divided into $G = 200$ intervals in order to reduce the error introduced by short-time approximation.}\label{fig11}
\end{figure}

The use of Trotter's formula, or the short time approximation improves the simulation fidelity significantly, as illustrated in Fig.~\ref{fig11}. For simulators with longitudinal or transverse Ising interactions, tuneable XY or Heisenberg interactions, we find that
the more time steps the total simulation time is divided into, the higher simulation fidelity is obtained, as in Figs.~\ref{fig11} (a), (b), (c), and (e). We believe that the error introduced by short time approximation can be reduced in this way. Nevertheless, for simulators with fixed-parameter XY or Heisenberg interaction Hamiltonians, we notice that the overall fidelity is sensitive to that of the sub-circuit used to simulate $U_{l,l+1}^{xx+yy}[\pi / (4 J_l)] = e^{\pm i\tau (\sigma_l^x \sigma_{l+1}^x + \sigma_l^y \sigma_{l+1}^y) \pi / 4}$ $(l = 1, \dots, N-1)$ (see Figs.~\ref{fig11} (d) and (f)). On the other hand, on the contrary to the error due to short-time approximation, the error from simulating $U_{l,l+1}^{xx+yy}[\pi / (4 J_l)]$ increases with the running times of the sub-circuit.
Figures~\ref{fig11}(d) and (f) show that when the simulation time is relatively short and the error from the Trotter's formula is small, the overall simulation fidelity is mainly determined by that of simulating $U_{l,l+1}^{xx+yy}[\pi / (4 J_l)]$. Therefore, we conclude that in this case the more time steps the total time for completing the whole task is divided into, the more time $U_{l,l+1}^{xx+yy}[\pi / (4 J_l)]$ simulation process is run, the larger the error is. When the simulation time becomes longer, error from theTrotter's formula becomes dominant. Therefore, as shown in Figs. ~\ref{fig11}(d) and (f), all curves are almost flat at the short simulation time and drop when the time increases. Moreover, the fidelity of simulation with smaller number $M$ of intervals is higher at shorter simulation time but lower at longer simulation time.

\begin{figure}
  \includegraphics[bb=108 212 470 545, width=8cm, clip]{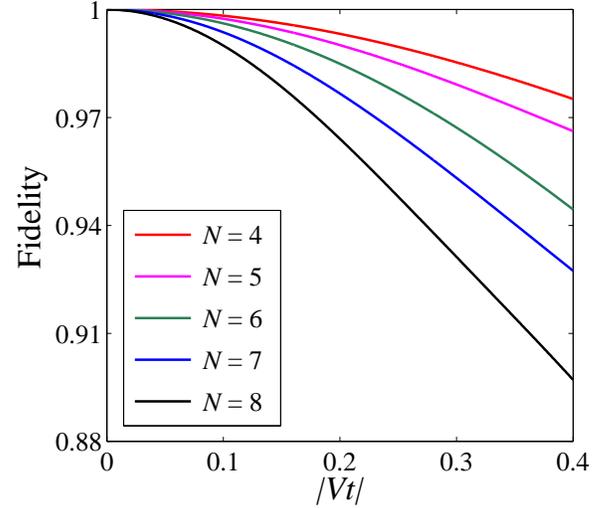}
  \caption{Fidelity of the algorithm run by quantum simulator with transverse Ising type nearest-neighbor interactions containing different number of qubits. Every term in the Hamiltonian of the simulator can be turned on and off during the simulation process. $\varepsilon_l = 2\times10^{12}$Hz, $\omega_l = 5\times10^9$Hz $(l = 1, \dots, N)$, $V_l = -2\times10^8$Hz, $J_l = 3\times10^7$Hz $(l = 1, \dots, N-1)$. The whole simulation task is divided into $M=20$ intervals.}
  \label{fig12}
\end{figure}

Figure~\ref{fig12} suggests that if other conditions are the same, the fidelity of simulation algorithms decreases when the qubit number in quantum simulators increases. In practice, this error can be compensated by reducing the error due to the short-time approximation.

\section{Complexity Analysis} \label{complex}

Let us now analyze complexity of the algorithms, which is determined by the total number of external single-qubit logic gate required for the simulation process. We will show that the algorithms are polynomial.

The number of single-qubit gates in any of quantum circuits in the previous sections increases linearly with the number of qubits $N$ in the simulator. One can extract $U_l^z(\tau) = \exp(-i\tau \omega_l \sigma_l^z / 2)$ in the complexity $O(N)$. The simulation of individual  $e^{-itH_{p0}} = \bigotimes_{m=1}^N U_m^z ( \varepsilon_m t/\omega_m )$ requires $O(N^2)$ external gates. We find that $U_{l,l+1}^{zz}(\tau) = \exp(-i\tau J_l \sigma_l^z \sigma_{l+1}^z)$ or $U_{l,l+1}^{xx+yy}(\tau) = \exp[-i J_l (\sigma_l^x \sigma_{l+1}^x + \sigma_l^y \sigma_{l+1}^y)]$ can also be simulated within $O(N)$. As shown in~\cite{Wu02},  it can be shown
\begin{equation}
e ^ {i\frac{\pi}{2} X_{l,l+1}} U_{m,l}^{zz}(\tau) e ^ {-i\frac{\pi}{2} X_{l,l+1}} = U_{m,l+1}^{zz}(\tau),
\end{equation}
for $m<l$, where $X_{l,m}=(\sigma^x_{l}\sigma^x_{m}+\sigma^y_{l}\sigma^y_{m})/2$. Thus it needs $O(N^2)$ to simulate an arbitrary long-range interaction $U_{m,l}^{zz}(\tau)$. Since $N(N+1) / 2$ terms are included in the pairing Hamiltonian Eq.~(\ref{BCS1}), simulation of all these interactions requires $O(N^4)$ logic gates, or the complexity of the whole algorithm is $O(N^4)$.

\begin{figure}[htb]
  \centerline{\includegraphics[width = 8 cm]{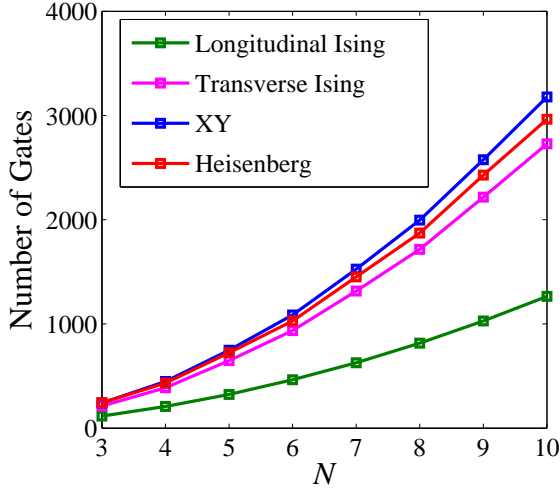}}
  \caption{
 Total number of external gates in the simulation process for four types of simulators with constant parameters. Here the complexity is shown for a single running process, in which both the simulation task and subprocess to simulate $e^{\pm i (\sigma_l^x \sigma_{l+1}^x + \sigma_l^y \sigma_{l+1}^y) \pi / 4}$ $(l = 1, \dots, N-1)$ are not further divided into any subintervals ($M=1$, $G=1$).
  }
  \label{fig13}
\end{figure}

\begin{figure}[htb]
  \centerline{\includegraphics[width = 8 cm]{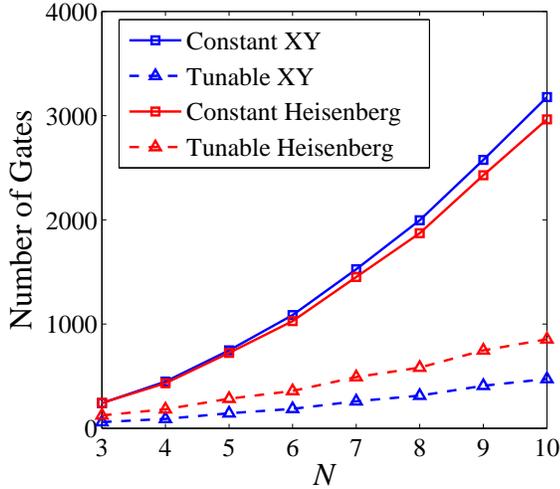}}
  \caption{
  Effects of tunable parameters on the complexity for XY type and Heisenberg type simulators. Solid lines stand for the case when all parameters in the simulator are constant, while dashed lines give the result when they are all tunable. Here the total number of external gates is shown for a single running process, in which both the simulation task and subprocess to simulate $e^{\pm i (\sigma_l^x \sigma_{l+1}^x + \sigma_l^y \sigma_{l+1}^y) \pi / 4}$ $(l = 1, \dots, N-1)$ are not further divided into any subintervals ($M=1$, $G=1$).
  }\label{fig14}
\end{figure}

In Fig.~\ref{fig13} we calculate total numbers of external gates for four types of simulators with all constant parameters when $N \leq 10$. It shows that longitudinal Ising simulator has the lowest complexity. On the other hand if the parameters of the simulators are tunable, the complexity is significantly reduced when the scale of the simulator increases, as shown in Fig.~\ref{fig14}.

\begin{figure}[htb]
  \centerline{\includegraphics[width = 8 cm]{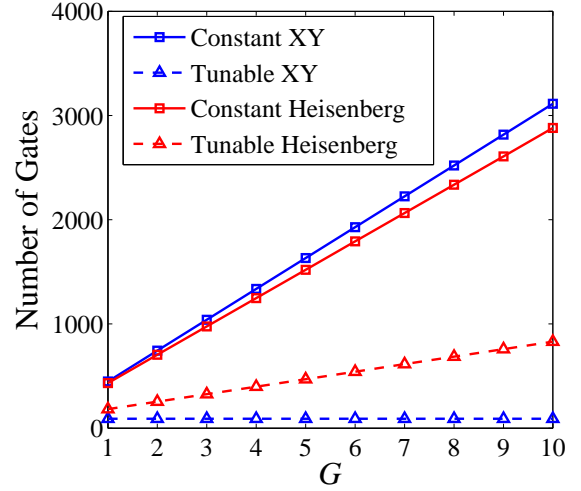}}
  \caption{
  Increasing of complexity due to more subintervals the processes for simulating $e^{\pm i (\sigma_l^x \sigma_{l+1}^x + \sigma_l^y \sigma_{l+1}^y) \pi / 4}$ $(l = 1, \dots, N-1)$ are divided into. XY and Heisenberg simulators are taken as the examples. Solid lines stand for the case when all parameters in the simulator are constant, while dashed lines give the result when they are all tunable. Here the whole simulation task is not further divided into any subintervals ($M=1$).
  }
  \label{fig15}
\end{figure}

Furthermore, although higher simulation fidelity can be obtained through the short-time approximation by dividing the simulation process into $M > 1$ intervals, this is at the expense of higher complexity, since the number of logic gates is $M$ times of the gate number without the short-time approximation. Moreover, although Trotter's formula can be applied to the subprocess of simulating $\exp[\pm i  \pi (\sigma_l^x \sigma_{l+1}^x + \sigma_l^y \sigma_{l+1}^y)/ 4]$ $(l = 1, \dots, N-1)$, the complexity grows linearly with the number $G$ of intervals that this subprocess is divided into, which is shown in Fig.~\ref{fig15}. We can also find that if simulators with tunable parameters are available, the effect of $G$ on the total number of external gates will be reduced. In particular, when all parameters of an XY type simulator are tunable, the growing of $G$ has no effect on the simulation complexity.

\section{Measurement Scheme} \label{section_mea}

\begin{figure}[htb]
  \centerline{\includegraphics[width = 8.5 cm]{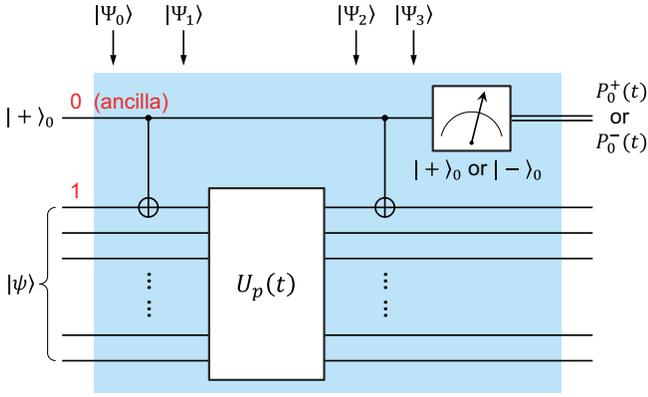}}
  \caption{Measurement scheme for the simulation algorithms in order to extract information such as energy spectrum from the simulator. ${|+ \rangle}_{0}$ and $|\psi \rangle$ stand for the initial state of the ancillary qubit and the quantum simulator respectively. Index 0 and 1 denote the acilla and the single qubit in the simulator directly coupled with the ancilla respectively. $|\Psi_0 \rangle$, $|\Psi_1 \rangle$, $|\Psi_2 \rangle$ and $|\Psi_3 \rangle$ are the states of the whole system during the simulation process. $U_p(t)$ is the time evolution operator including the complete simulation algorithms, which is explicitly equivalent to the pairing Hamiltonian. Two CNOT gates in which qubit 0 (ancilla) serves as the control qubit and qubit 1 serves as the target qubit are graphically shown. Measurement is done to the ancilla on ${|+ \rangle}_{0}$ or ${|- \rangle}_{0}$ basis, giving probability $P_0^+(t)$ or $P_0^-(t)$ as the result respectively, where $t$ corresponds to the evolution time for BCS system in $U_p(t)$.}
  \label{fig16}
\end{figure}

Now we come to the measurement or readout approach as the last step of our simulation algorithm.  The measurement circuit is shown in Fig.~\ref{fig16} directly after the simulation for paring Hamiltonian. We use an ancillary qubit, denoted by a red color $0$ and entangled with the simulator. We measure the ancilla at the end of the simulation process. We use one qubit, marked by a red color $1$, to directly interact with the ancillary qubit in the whole simulator. ${|\pm \rangle}_{0} = ({|\uparrow\: \rangle}_{0} \pm {|\downarrow\: \rangle}_{0})/\sqrt{2}$ are eigenstates of operator $\sigma_x^0$ for the ancillary qubit.

The readout processes as follows. First, the ancillary qubit is prepared to the state ${|+ \rangle}_{0}$, while the simulator, including qubits $1$ to $N$ involved in the algorithms, is prepared to a state $|\psi \rangle$. The whole system is initially at $|\Psi_0 \rangle={|+ \rangle}_{0}\otimes |\psi \rangle$. We then apply a controlled-NOT gate denoted by $\bf{CNOT_{0 \rightarrow 1}}$  to qubit $0$ and qubit $1$ so that the system state becomes $|\Psi_1 \rangle = {\bf{CNOT_{0 \rightarrow 1}}} |\Psi_0 \rangle$, where the ancilla serves as the control qubit and the qubit $1$ serves as the target qubit. We run the complete simulation algorithms for a time interval $t$, represented by $U_{p}(t) = e^{-itH_{p}}$ such that the system becomes $|\Psi_2 \rangle = \left[ I_0 \otimes U_{p}(t) \right] |\Psi_1 \rangle$. A new controlled-NOT gate is applied to the qubit $0$ and the qubit $1$ again, where the ancilla again serves as the control qubit and the qubit $1$ serves as the target qubit.  The state of the whole system ends up with $|\Psi_3 \rangle = {\bf{CNOT_{0 \rightarrow 1}}} |\Psi_2 \rangle$.
Finally, the ancilla is measured on $\left\{ {|+ \rangle}_0, {|- \rangle}_0 \right\}$ basis. The probabilities of obtaining the states $|\pm\rangle$ are $P_0^\pm(t)$, respectively, which vary with time $t$. We can use either $P_0^+(t)$ and $P_0^-(t)$  to extract spectrum information of the simulator. For example, $P_0^+(t)$ is calculated as
\begin{equation}
P_0^+(t) = \frac{1}{2} + \frac{1}{4} \left[\langle \psi | \sigma_1^x(t) \sigma_1^x(0) |\psi \rangle + \textrm{c.c.}\right]\label{prob},
\end{equation}
where $\sigma_1^x (t) = U_{p}^\dagger (t) \sigma_1^x U_{p}(t)$ is in the Heisenberg picture.

We could now expand the initial state with a complete set of quantum numbers $n$, $i$ and $\beta_i$, $
|\psi\rangle=\sum_{n, i, \beta_i} B_{n, i, \beta_i} \left|n,i,\beta_i\right\rangle$. Here $n$ is the total number of spin-up qubits in our simulator. $i$ denotes the energy level $E_{n,i}$ with a given $n$ and $\beta_i$ denotes the degeneracy for a given energy level $E_{n,i}$. Physically, $n$ is also the number of Cooper pairs in the simulated pairing Hamiltonian. States with different quantum number $n$ are mutually orthogonal, $\langle m,j,\beta_j | n,i,\beta_i \rangle = \delta_{nm} \langle m,j,\beta_j | n,i,\beta_i \rangle$. In order to simplify the calculation, we can specially prepare qubit $1$ on the spin-up state, such that $\sigma_1^z |\psi \rangle = |\psi \rangle$ and
\begin{equation} \label{colxx}
\begin{split}
\langle \psi | \sigma^x_1(t)\sigma^x_1(0) |\psi \rangle =
\sum_{n,i,j} \tilde{C}_{n, i, j}  e^{it\left(E_{n,j} - E_{n-1, i}\right)}
\end{split}
\end{equation}
where $\tilde{C}_{n, i, j} = \sum_{\beta_i, \beta_j}B^*_{n, j, \beta_j} B_{n, i, \beta_i} \left\langle n,j,\beta_j|n,i,\beta_i \right\rangle$.  To study the measurement result in frequency domain, we can take the Fourier transform  $\tilde{\rho}_0^+ (\omega)$ of $P_0^+(t)$,
\begin{equation}
\begin{split}
\tilde{\rho}_0^+ (\omega)
& = \pi \delta(\omega) \\
& \quad \, + \frac{\pi}{2} \sum_{n, i, j} \tilde{C}_{n, i, j} \delta \left(\omega + E_{n,j} - E_{n-1, i} \right) \\
& \quad \, + \frac{\pi}{2} \sum_{n, i, j} \tilde{C}_{n, i, j}^* \delta \left(\omega -E_{n,j} + E_{n-1, i}\right).
\end{split}
\end{equation}
The spectrum of $P_0^+(t)$ is symmetric on amplitude and antisymmetric on phase because $P_0^+(t)$ is real. In amplitude spectrum, there should be sharp peaks, ideally $\delta$-shape peaks at frequencies $\omega_{n,i,j}^\pm = \pm \left( E_{n,j} - E_{n-1, i}\right)$.  Energy spectra of the pairing Hamiltonian could be extracted from these frequencies. For instance, the energy gap between the ground state and the first excitation state with $n$-Cooper pairs can be obtained by
\begin{eqnarray}
2\Delta_n &=& E_{n, 1} - E_{n, 0} = \left(E_{n, 1} - E_{n-1, i}\right) - \left(E_{n, 0} - E_{n-1, i}\right) \nonumber\\
&=& \omega_{n,i,1}^+ - \omega_{n,i,0}^+.
\end{eqnarray}

Generally, low-lying energy spectra of a pairing Hamiltonian are of great interest. Using the adiabatic method developed in Ref. \cite{Wu02}, we can prepare initial states that involve  a few eigenstates of the simulated pairing Hamiltonian to improve the efficiency of our measurement approach.

\section{Conclusion} \label{conclusion}

In summary, we study algorithms for simulating the pairing Hamiltonians based on various nearest-neighbor interactions, e.g., longitudinal Ising Hamiltonian, transverse Ising Hamiltonian, $XY$ Hamiltonian and the Heisenberg Hamiltonian, which are available in the solid-state quantum devices. With current experimental advances on gate fidelity and coherence time, our proposal might be feasible in superconducting qubit circuits. Since the four types of nearest-neighbor interactions are shared by various quantum systems~\cite{Loss98, Kane98, Vrijen00}, our algorithms are adaptable to various solid-state systems.

\end{document}